\documentclass[aip,jcp,groupedaddress]{revtex4-2}

\usepackage{amsfonts}
\usepackage{graphicx}
\usepackage{amsmath}
\usepackage{color}
\usepackage{float}

\begin{document}

\title{Nuclear Spin Isomers and the Pauli Principle in Polaritonic Chemistry}

\author{Csaba F\'abri}
\email{fabri.csaba@science.unideb.hu}
\affiliation{Department of Theoretical Physics, University of Debrecen, P.O. Box 400, H-4002 Debrecen, Hungary}

\author{G\'abor J. Hal\'asz}
\affiliation{Department of Information Technology, University of Debrecen, P.O. Box 400, H-4002 Debrecen, Hungary}

\author{Lorenz S. Cederbaum}
\affiliation{Theoretische Chemie, Physikalisch-Chemisches Institut, Universit\"at Heidelberg, D-69120 Heidelberg, Germany}

\author{\'Agnes Vib\'ok}
\email{vibok@phys.unideb.hu}
\affiliation{Department of Theoretical Physics, University of Debrecen, P.O. Box 400, H-4002 Debrecen, Hungary}
\affiliation{ELI-ALPS, ELI-HU Non-Profit Ltd, H-6720 Szeged, Dugonics t\'er 13, Hungary}

\date{\today}

\begin{abstract}
The Pauli principle has far-reaching consequences in quantum physics. 
Here, we investigate, for the first time, its implications, together with nuclear spin isomerism, in polaritonic chemistry.
The theory is developed for a single and a few molecules as well as for an ensemble of molecules. As an explicit and detailed example
we first present an accurate numerical description of a realistic situation involving two $^{14}$NH$_3$ molecules, existing as ortho and para spin isomers, in an infrared plasmonic cavity. 
Then, we generalize the approach for molecular ensembles using analytical considerations based on the Tavis--Cummings model and
simulate the transmission spectrum of a Fabry--P\'erot cavity filled with $^{14}$NH$_3$ gas using quantum mechanics.
These results are directly relevant for recent gas-phase experiments studying rovibrational polaritons in molecules.
Our findings undoubtedly demonstrate that the Pauli principle and nuclear spin isomerism significantly reshape collective light–matter coupling involving molecules with identical nuclei. 
\end{abstract}

\maketitle

\section{Introduction}

Polaritonic chemistry has emerged as a novel and powerful paradigm for modifying the energy landscape of molecules over the past fifteen years.
Strong coupling between molecules and confined electromagnetic modes gives rise to hybrid light-matter states, called polaritons.
By hybridizing molecular excitations with cavity photons, 
one can alter the rate of chemical reactions \cite{12HuScGe,16ThGeSh} and energy transfer processes \cite{16ZhChWa} as well as the dynamical and spectroscopic properties of molecules,
\cite{15GaGaFe,16KoBeMu,17FlRuAp,17HeSp_2,17LuFeTo,18GrTo,18Vendrell,18Vendrell_2,18PiScCh,19CaRiZh,19GaClGa,20TaMaZh,20HaRoKj,20LiSuNi,21AhHuBe,21LiNiSu,21LiMaHu,21MeGaFe,21CeKu,22ReSoGe,22CsVeHa,22DuYu,23ScSiRu,23ZePeEc,23ScKo_2,24FeShRo,25CsSzVi,25CaLeKo,25SiHo,25MoVaCu,25MoVaNi,25KoYu,26AkShCo}
including light-induced nonadiabatic effects,\cite{18SzHaCs_2,21FaHaCe,24FaCsHa_2}
and also enhance exciton and charge transport.\cite{15ScGeTi,17HaScSc}

Despite remarkable progress in the field of polaritonic chemistry, surveyed in an array of excellent review articles \cite{18RiMaDu,21FeFeGa,22FrGaFe,22LiCuSu,23FrCo,23MaTaWe,24BoWe,25SiGaLa} and in a book,\cite{25YuGiRi}
the origin of certain cavity-induced effects, such as the modification of reaction rates, is still not fully understood.
In addition, a notable conceptual gap exists between polaritonic chemistry and more mature areas of physical chemistry. For example, in molecular spectroscopy and reaction dynamics, the Pauli principle and the exchange symmetry of identical particles play a pivotal role, giving rise to nuclear spin statistics \cite{06BuJe} and symmetry selection rules.\cite{77Quack}
It is also well established that molecules having identical nuclei can exist in different nuclear spin isomers which can convert into each other under appropriate conditions.\cite{05HoOk}
However, the Pauli principle and nuclear spin degrees of freedom seem largely overlooked in polaritonic chemistry.
It is therefore essential to bring well-established exchange symmetry principles into polaritonic chemistry in order to obtain a correct quantum-mechanical description of strong coupling. 

\begin{figure}[ht]
\includegraphics[scale=0.5]{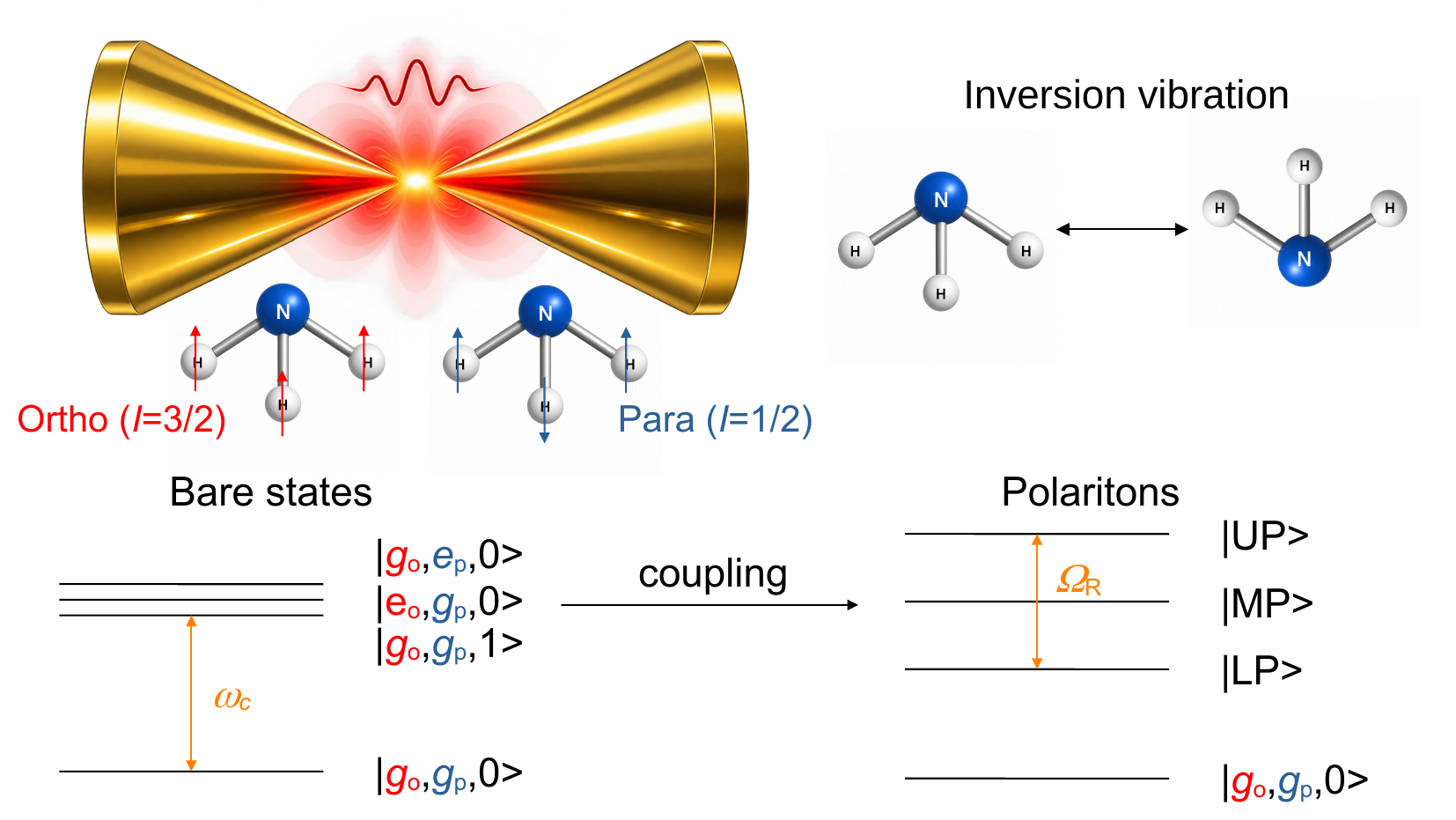}
\caption{\label{fig:2mol_illustration}
Illustration of an ortho and a para spin isomer of $^{14}$NH$_3$ coupled to a plasmonic cavity. 
The cavity frequency $\omega_\textrm{c}$ is tuned to resonance with a rovibrational transition involving the inversion vibrational mode.
Bare states in the so-called singly-excited manifold are (nearly) degenerate.
Here, the lower and upper eigenstates resonantly coupled by the cavity mode are denoted by
$|g_\textrm{o} \rangle$/$|e_\textrm{o} \rangle$ (ortho isomer)
and $|g_\textrm{p} \rangle$/$|e_\textrm{p} \rangle$ (para isomer), while $|0\rangle$ and $|1\rangle$ refer to Fock states of the cavity mode with 0 and 1 photons.
Due to cavity-molecule coupling, rovibrational polaritons (LP: lower polariton, MP: middle polariton, UP: upper polariton) are formed.}
\end{figure}

\begin{figure}[ht]
\includegraphics[scale=0.55]{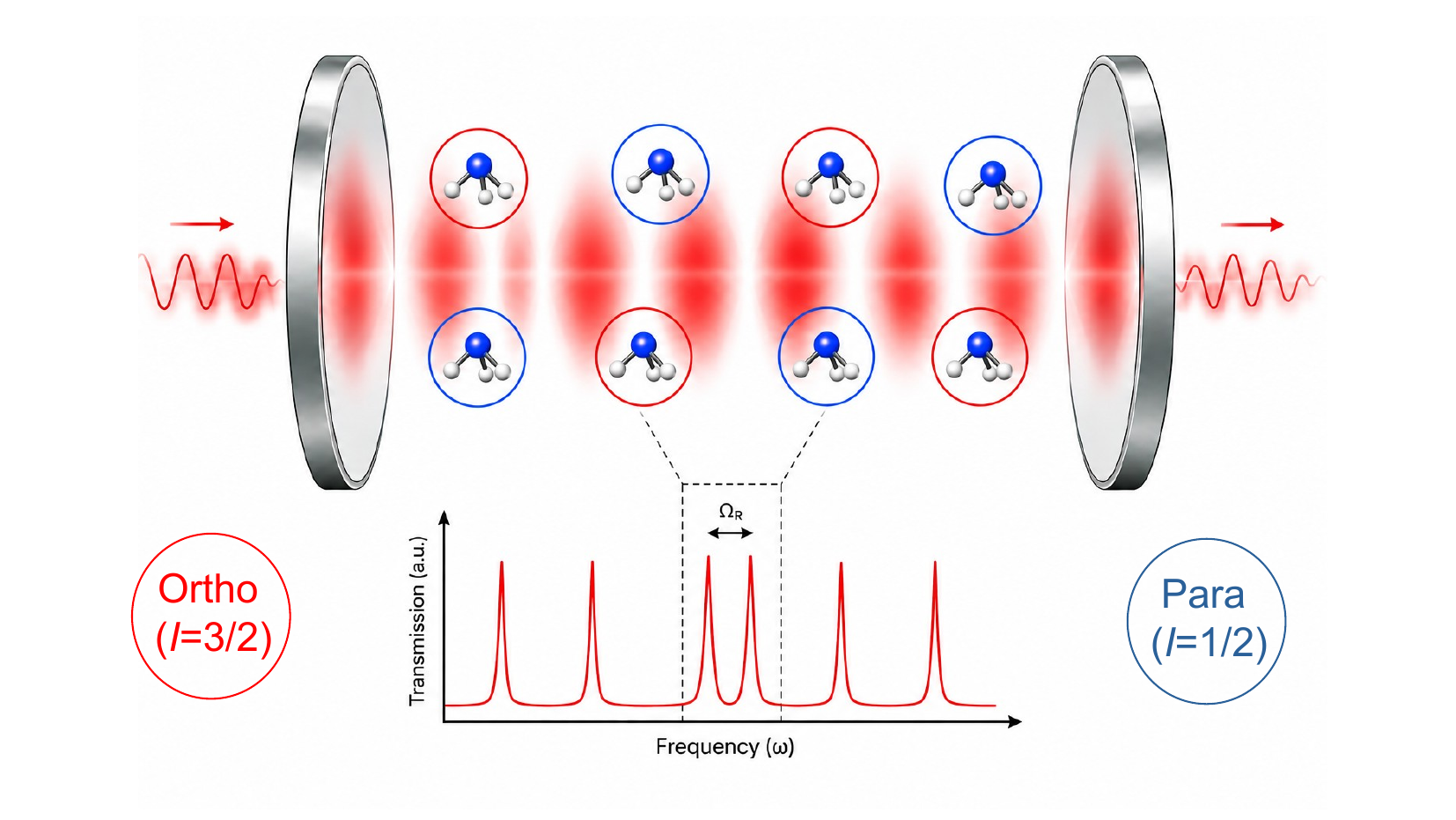}
\caption{\label{fig:nmol_illustration}
Fabry--P\'erot cavity coupled to an ensemble of ortho (shown in red) and para (shown in blue) spin isomers of $^{14}$NH$_3$.
The cavity couples to the inversion vibration and the cavity transmission spectrum is recorded.
Resonant cavity-molecule coupling results in the splitting of the resonance in the middle of the cavity transmission spectrum.
As demonstrated later, the resonance highlighted in the figure can be split into two or three peaks, depending on whether only one or both spin isomers can couple to the cavity mode.}
\end{figure}

To the best of our knowledge, only a recent study \cite{23Szidarovszky} has explicitly addressed the exchange of identical molecules in connection with the Pauli principle in molecular polaritons. 
We also mention that the exchange of molecules was studied concerning the generalization of the Tavis--Cummings model for multi-level anharmonic systems.\cite{21CaRiYu}
While Ref. \citenum{23Szidarovszky} provides a symmetry analysis of the accessible state space for an ensemble of identical molecules, it does not explore how exchange symmetry manifests in experimentally relevant scenarios.
In this work, we intend to close this gap by incorporating nuclear spin degrees of freedom and exchange symmetry of identical nuclei into the description of molecular polaritons.
We demonstrate that the Pauli principle and nuclear spin isomerism have a striking impact on the quantum dynamics and spectroscopy of coupled cavity-molecule systems and also reshape collective light–matter coupling.
To this end, we apply a full-dimensional rotational-vibrational model of ammonia ($^{14}$NH$_3$) which exists in two distinct nuclear spin isomers, ortho and para.
First, we focus on the highly accurate numerical description of two non-interacting $^{14}$NH$_3$ molecules coupled to an infrared (IR) plasmonic cavity (see Fig. \ref{fig:2mol_illustration}).
Then, we employ the Tavis--Cummings model \cite{68TaCu} to study effects related to the Pauli principle and nuclear spin isomers in a molecular ensemble coupled to a Fabry--P\'erot cavity operating in the IR range (see Fig. \ref{fig:nmol_illustration}).
Namely, we consider a Fabry--P\'erot cavity filled with a large number of $^{14}$NH$_3$ molecules and simulate the cavity transmission spectrum using a full quantum description.
In both cases, vibrational strong coupling (VSC) is assumed.
We demonstrate that for certain cavity parameters, both spin isomers can participate in the formation of collective molecular polaritons.
In other cases, only one spin isomer can couple to the cavity, while the other spin isomer acts as a spectator.

Before moving forward, a few comments are in order.
In Fabry--Pérot cavities, single-molecule VSC is severely constrained by large mode volumes imposed by the IR wavelength.
Nevertheless, VSC was achieved experimentally in molecular ensembles both in condensed \cite{14LoSi,15ShGeHu,16ThGeSh,16DuSpFe} and gas phases.\cite{23WrNeWe,23WrNeWe_2,24NeWe}
Plasmonic nanocavities \cite{18HuSiHu} provide subwavelength confinement, leading to significantly enhanced electric field strengths, but low quality factors generally preclude single-molecule VSC.
However, single-molecule electronic strong coupling was realized for dye molecules \cite{16ChNiBe} and quantum dots \cite{16SaBiCh} experimentally.
In the case of two $^{14}$NH$_3$ molecules, we assume a plasmonic nanocavity setup and identify striking effects related to the Pauli principle in polaritonic chemistry.
Our findings for the two-molecule case are expected to persist or even become more pronounced for larger numbers of molecules for which VSC can be attained by current experimental approaches.
This statement is validated by studying a molecular ensemble coupled to a Fabry--P\'erot IR cavity using experimentally feasible cavity parameters.
The ensemble results also highlight the importance of nuclear spin degrees of freedom and are of direct relevance for recent gas-phase polaritonic experiments.\cite{23WrNeWe,23WrNeWe_2,24NeWe}
We stress that the effects identified in this work are generally expected in molecules with identical nuclei.

\section{General theory}

In this section, we first introduce the Hamiltonian used to describe the coupled cavity-molecule system in accurate numerical computations for two molecules.
Then, we generalize the Tavis--Cummings model for the case of a molecular ensemble consisting of two distinct nuclear spin isomers.
Finally, we describe the quantum-mechanical framework used to calculate cavity transmission spectra.

\subsection{Cavity-molecule Hamiltonian}

The coupled cavity-molecule system is described by the Pauli--Fierz Hamiltonian \cite{97CoDuGr,23MaTaWe}
\begin{equation}
        \hat{H}_\textrm{cm} = \hat{H}_\textrm{m} + \hbar \omega_\textrm{c} \hat{a}^\dagger \hat{a} - g \hat{\vec{\mu}} \vec{e} (\hat{a}^\dagger + \hat{a}) +
            \frac{g^2}{\hbar \omega_\textrm{c}} (\hat{\vec{\mu}} \vec{e})^2
   \label{eq:Hcm}
\end{equation}
where $\hat{H}_\textrm{m}$ is the Hamiltonian of the isolated (field-free) molecule, 
$\omega_\textrm{c}$ denotes the angular frequency of the cavity mode, 
$\hat{a}^\dagger$ and $\hat{a}$ are creation and annihilation operators of the cavity mode, 
$\hat{\vec{\mu}}$ is the electric dipole moment operator of the molecule and 
$\vec{e}$ corresponds to the polarization vector of the cavity field. 
The cavity-molecule coupling is characterized by the coupling strength parameter 
$g = \sqrt{\frac{\hbar \omega_\textrm{c}}{2 \epsilon_0 V}}$
where $\epsilon_0$ and $V$ are the vacuum permittivity and the cavity mode volume, respectively.
The last term of $\hat{H}$ is the so-called dipole self-energy (DSE) whose significance has been examined thoroughly.\cite{18RoWeRu,20ScRuRo}

Next, the Hamiltonian of Eq. \eqref{eq:Hcm} is generalized to $n$ molecules, which gives
\begin{gather}
        \hat{H} = \sum_{i=1}^n \hat{H}_0^{(i)} + \hbar \omega_\textrm{c} \hat{a}^\dagger \hat{a}
        - g \sum_{i=1}^n \vec{\mu}_0^{(i)} \vec{e} (\hat{a}^\dagger + \hat{a}) - 
        \frac{g^2}{2} \sum_{i=1}^n \vec{e} \alpha_0^{(i)} \vec{e} (\hat{a}^\dagger + \hat{a})^2 \nonumber \\
        + \frac{g^2}{\hbar \omega_\textrm{c}} \Biggl[ \sum_{i=1}^n \vec{e} d_0^{(i)} \vec{e} + \sum_{i=1}^n \sum_{j \ne i}^n \left( \vec{\mu}_0^{(i)} \vec{e} \right) \left( \vec{\mu}_0^{(j)} \vec{e} \right) \Biggr]. \label{eq:Hcm_gs_mult}
\end{gather}
Here, the treatment is restricted to the electronic ground state, light-matter interaction terms are included up to $g^2$ and intermolecular interactions are neglected
(see the Supporting Information for more information).
In Eq. \eqref{eq:Hcm_gs_mult}, $\hat{H}_0^{(i)} = \hat{T}^{(i)} + V_0^{(i)}$ is the ground-state molecular Hamiltonian where $\hat{T}^{(i)}$ and $V_0^{(i)}$
equal the kinetic energy operator of the nuclei and the ground-state potential energy surface of the $i$th molecule, respectively.
Furthermore, $\vec{\mu}_0^{(i)} = \langle \varphi_0^{(i)} | \hat{\vec{\mu}}^{(i)} | \varphi_0^{(i)} \rangle$\
is the dipole moment vector in the electronic ground state $| \varphi_0^{(i)} \rangle$,
$\alpha_0^{(i)}$ denotes the ground-state polarizability tensor and
$(d_0)^{(i)}_{jk} = \langle \varphi_0^{(i)} | \hat{\mu}_j^{(i)} \hat{\mu}_k^{(i)} | \varphi_0^{(i)} \rangle$
refers to the components ($j,k=1,2,3$) of the ground-state DSE tensor.

\subsection{Generalization of the Tavis--Cummings model}
\label{sec:tc}

Next, we generalize the Tavis--Cummings model \cite{68TaCu} for a molecular ensemble consisting of two different (labeled as ortho and para) spin isomers.
The derivations presented in this section provide interesting insight into the formation of collective ortho-para polaritons and will be used extensively to interpret our results.
Let us consider $n_\textrm{o}$ ortho and $n_\textrm{p}$ para molecules and couple a pair of close-lying ortho and para rovibrational transitions to an IR cavity mode.
In this case, the basis of the singly-excited manifold involving molecular eigenstates coupled by the cavity can be constructed as
\begin{gather}
    | b \rangle = |g_\textrm{o}^{(1)},\dots,g_\textrm{o}^{(n_\textrm{o})},g_\textrm{p}^{(1)},\dots,g_\textrm{p}^{(n_\textrm{p})},1\rangle \nonumber \\
    | b_\textrm{o}^{(i)} \rangle = |g_\textrm{o}^{(1)},\dots,e_\textrm{o}^{(i)},\dots,g_\textrm{o}^{(n_\textrm{o})},g_\textrm{p}^{(1)},\dots,g_\textrm{p}^{(n_\textrm{p})},0\rangle \label{eq:TC_basis} \\
    | b_\textrm{p}^{(j)} \rangle = |g_\textrm{o}^{(1)},\dots,g_\textrm{o}^{(n_\textrm{o})},g_\textrm{p}^{(1)},\dots,e_\textrm{p}^{(j)},\dots,g_\textrm{p}^{(n_\textrm{p})},0\rangle \nonumber
\end{gather}
where $i = 1,\dots,n_\textrm{o}$, $j = 1,\dots,n_\textrm{p}$,
$|g_\textrm{o}^{(i)}\rangle/|e_\textrm{o}^{(i)}\rangle$ and
$|g_\textrm{p}^{(j)}\rangle/|e_\textrm{p}^{(j)}\rangle$
denote ortho and para lower and upper eigenstates of the selected ortho and para transitions, respectively, and the last number in the kets of Eq. \eqref{eq:TC_basis} refers to Fock states of the cavity mode.

Next, we neglect the DSE and polarizability terms both in theoretical considerations and numerical computations for the molecular ensemble.
Under these assumptions, the matrix of the cavity-molecule Hamiltonian is constructed in the singly-excited manifold, which yields the so-called arrowhead matrix \cite{84WaCeSc}
\begin{equation}
    \mathbf{H} = 
    \left[
    \begin{array}{c|ccc|ccc}
            0 & V_\textrm{o}^{(1)} & \dots & V_{\textrm{o}}^{(n_\textrm{o})} & V_{\textrm{p}}^{(1)} & \dots & V_{\textrm{p}}^{(n_\textrm{p})} \\
            \hline
            V_{\textrm{o}}^{(1)} & \Delta_\textrm{o} & \dots & 0 & 0 & \dots & 0 \\
            \vdots & \vdots & \ddots & \vdots & \vdots & \ddots & \vdots \\
            V_{\textrm{o}}^{(n_\textrm{o})} & 0 & \dots & \Delta_\textrm{o} & 0 & \dots & 0 \\
            \hline
            V_{\textrm{p}}^{(1)} & 0 & \dots & 0 & \Delta_\textrm{p} & \dots & 0 \\
            \vdots & \vdots & \ddots & \vdots & \vdots & \ddots & \vdots \\
            V_{\textrm{p}}^{(n_\textrm{p})} & 0 & \dots & 0 & 0 & \dots & \Delta_\textrm{p}
        \end{array}
        \right].
    \label{eq:arrowhead}
\end{equation}
In Eq. \eqref{eq:arrowhead}, the ortho and para detunings are defined as 
$\Delta_\textrm{o} = E_\textrm{o}^{(e)}-E_\textrm{o}^{(g)}-\hbar\omega_\textrm{c}$ and
$\Delta_\textrm{p} = E_\textrm{p}^{(e)}-E_\textrm{p}^{(g)}-\hbar\omega_\textrm{c}$,
where $E_\textrm{o}^{(g)}/E_\textrm{o}^{(e)}$ and
$E_\textrm{p}^{(g)}/E_\textrm{p}^{(e)}$
denote ortho and para lower and upper energy levels, respectively.
In Eq. \eqref{eq:arrowhead}, the energy scale is defined by setting the quantity $n_\textrm{o} E_\textrm{o}^{(g)}+n_\textrm{p}E_\textrm{p}^{(g)}+\hbar\omega_\textrm{c}$ to zero.
In addition, the ortho and para dipole coupling matrix elements are defined as
$V_\textrm{o}^{(i)} = -g \langle g_\textrm{o}^{(i)} | \vec{\mu}_0^{(i)} \vec{e}  | e_\textrm{o}^{(i)} \rangle$ and
$V_\textrm{p}^{(j)} = -g \langle g_\textrm{p}^{(j)} | \vec{\mu}_0^{(j)} \vec{e} | e_\textrm{p}^{(j)} \rangle$, respectively.

As shown in the Supporting Information, $\mathbf{H}$ has $n_\textrm{o}-1$ and $n_\textrm{p}-1$ dark eigenstates with degenerate eigenvalues $\Delta_\textrm{o}$ and $\Delta_\textrm{p}$, respectively.
Common to these $n_\textrm{o}+n_\textrm{p}-2$ dark states is that they give zero overlap with basis state $| b \rangle$.
In addition, $\mathbf{H}$ has three bright states whose energies can be computed by diagonalizing the three-dimensional bright Hamiltonian (see the Supporting Information)
\begin{equation}
    \mathbf{H}_\textrm{bright} = 
    \left[
    \begin{array}{ccc}
            0 & v_\textrm{o} & v_\textrm{p} \\
            v_\textrm{o} & \Delta_\textrm{o} & 0 \\
            v_\textrm{p} & 0 & \Delta_\textrm{p} 
        \end{array}
        \right]
    = \left[
    \begin{array}{ccc}
            0 & v_\textrm{o} & v_\textrm{p} \\
            v_\textrm{o} & \Delta+\delta & 0 \\
            v_\textrm{p} & 0 & \Delta-\delta
        \end{array}
        \right]
    \label{eq:mx_bright}
\end{equation}
where
$\Delta = (\Delta_\textrm{o}+\Delta_\textrm{p})/2$,
$\delta = (\Delta_\textrm{o}-\Delta_\textrm{p})/2$,
$v_\textrm{o} = \sqrt{\sum_{i=1}^{n_\textrm{o}}|V_\textrm{o}^{(i)}|^2}$
and
$v_\textrm{p} = \sqrt{\sum_{j=1}^{n_\textrm{p}}|V_\textrm{p}^{(j)}|^2}$.
In Eq. \eqref{eq:mx_bright}, the three-dimensional bright basis
\begin{gather}
    |w_0 \rangle = | b \rangle \nonumber \\
    |w_1 \rangle = \frac{1}{v_\textrm{o}} \sum_{i=1}^{n_\textrm{o}} V_\textrm{o}^{(i)} | b_\textrm{o}^{(i)} \rangle \label{eq:bright_basis} \\
    |w_2 \rangle = \frac{1}{v_\textrm{p}} \sum_{j=1}^{n_\textrm{p}} V_\textrm{p}^{(j)} | b_\textrm{p}^{(j)} \rangle \nonumber
\end{gather}
has been used to transform $\mathbf{H}$ of Eq. \eqref{eq:arrowhead} to $\mathbf{H}_\textrm{bright}$.
It is also shown in the Supporting Information that for $\delta = 0$ ($\Delta_\textrm{o} = \Delta_\textrm{p} = \Delta$), the eigenstate of $\mathbf{H}_\textrm{bright}$ with eigenvalue $\Delta$ becomes dark, while the remaining two bright eigenstates correspond to energies
$(\Delta \pm \sqrt{\Delta^2+4(v_\textrm{o}^2 + v_\textrm{p}^2)})/2$.

Molecular rovibrational eigenstates are often degenerate, which calls for generalization of the previous considerations based on the Tavis--Cummings model.
Let us assume that eigenstates involved in cavity coupling are degenerate and introduce additional labels to distinguish degenerate states.
Our updated notation for degenerate ortho and para ground and excited eigenstates reads
$|g_{\textrm{o},\kappa_i}^{(i)} \rangle$ ($\kappa_i = 1, \dots, d_\textrm{o}^g$),
$|e_{\textrm{o},\nu_i}^{(i)} \rangle$ ($\nu_i = 1, \dots, d_\textrm{o}^e$),
$|g_{\textrm{p},\lambda_j}^{(j)} \rangle$ ($\lambda_j = 1, \dots, d_\textrm{p}^g$)
and
$|e_{\textrm{p},\sigma_j}^{(j)} \rangle$ ($\sigma_j = 1, \dots, d_\textrm{p}^e$)
for the $i$th molecule.
Here, the quantities $d_\textrm{o}^g$, $d_\textrm{o}^e$, $d_\textrm{p}^g$ and $d_\textrm{p}^e$ denote degrees of degeneracy of the respective states.
In addition, we stipulate that dipole interaction matrix elements satisfy the relations
\begin{gather}
    \langle g_{\textrm{o},\kappa_i}^{(i)} | \vec{\mu}_0^{(i)} \vec{e} | e_{\textrm{o},\nu_i}^{(i)} \rangle = 
    \langle g_{\textrm{o},\kappa_i}^{(i)} | \vec{\mu}_0^{(i)} \vec{e} | e_{\textrm{o},\kappa_i}^{(i)} \rangle \delta_{\kappa_i \nu_i}\nonumber \\
    \langle g_{\textrm{p},\lambda_j}^{(j)} | \vec{\mu}_0^{(j)} \vec{e} | e_{\textrm{p},\sigma_j}^{(j)} \rangle =
    \langle g_{\textrm{p},\lambda_j}^{(j)} | \vec{\mu}_0^{(j)} \vec{e} | e_{\textrm{p},\lambda_j}^{(j)} \rangle \delta_{\lambda_j \sigma_j}.
\label{eq:dip_deg}
\end{gather}
Note that $d_\textrm{o}^g \ne d_\textrm{o}^e$ and $d_\textrm{p}^g \ne d_\textrm{p}^e$ in most cases.
We will show in Section \ref{sec:model} that in our case molecular eigenstates are degenerate in terms of the nuclear spin ($I$ and $M_I$) and $M$ (angular momentum projection along the space-fixed $z$ axis) quantum numbers.
Moreover, dipole matrix elements are diagonal with respect to both the nuclear spin and $M$ quantum numbers, which is a special case of Eq. \eqref{eq:dip_deg} with $\kappa_i$, $\nu_i$, $\lambda_j$ and $\sigma_j$ corresponding to the nuclear spin and $M$ quantum numbers.
Due to degeneracy, the basis of the singly-excited subspace defined in Eq. \eqref{eq:TC_basis} must be generalized.
First, let us define
\begin{equation}
     | b_{\kappa,\lambda} \rangle = 
     |g_{\textrm{o},\kappa_1}^{(1)},\dots,g_{\textrm{o},\kappa_{n_\textrm{o}}}^{(n_\textrm{o})},g_{\textrm{p},\lambda_1}^{(1)},\dots,g_{\textrm{p},\lambda_{n_\textrm{p}}}^{(n_\textrm{p})},1\rangle
\label{eq:TC_basis_deg_1}
\end{equation}
which, due to Eq. \eqref{eq:dip_deg} and the orthogonality of degenerate eigenstates, can be coupled by the dipole interaction only to basis functions
\begin{align}
   & | b_{\textrm{o},\kappa,\lambda}^{(i)} \rangle =
   |g_{\textrm{o},\kappa_1}^{(1)},\dots,e_{\textrm{o},\kappa_i}^{(i)},\dots,g_{\textrm{o},\kappa_{n_\textrm{o}}}^{(n_\textrm{o})},g_{\textrm{p},\lambda_1}^{(1)},\dots,g_{\textrm{p},\lambda_{n_\textrm{p}}}^{(n_\textrm{p})},0\rangle \nonumber \\
   & | b_{\textrm{p},\kappa,\lambda}^{(j)} \rangle =
   |g_{\textrm{o},\kappa_1}^{(1)},\dots,g_{\textrm{o},\kappa_{n_\textrm{o}}}^{(n_\textrm{o})},g_{\textrm{p},\lambda_1}^{(1)},\dots,e_{\textrm{p},\lambda_j}^{(j)},\dots,g_{\textrm{p},\lambda_{n_\textrm{p}}}^{(n_\textrm{p})},0\rangle.
\label{eq:TC_basis_deg_2}
\end{align}
The state $| b_{\kappa,\lambda} \rangle$ can be constructed in 
$(d_\textrm{o}^{g})^{n_\textrm{o}} (d_\textrm{p}^{g})^{n_\textrm{p}}$ ways.
In addition, notations introduced for the bright basis are also updated to account for degeneracies, that is,
\begin{gather}
    |w_{0,\kappa,\lambda} \rangle = | b_{\kappa,\lambda} \rangle \nonumber \\
    |w_{1,\kappa,\lambda} \rangle = \frac{1}{v_{\textrm{o},\kappa}} \sum_{i=1}^{n_\textrm{o}} V_{\textrm{o},\kappa_i}^{(i)} | b_{\textrm{o},\kappa,\lambda}^{(i)} \rangle \label{eq:bright_basis_deg} \\
    |w_{2,\kappa,\lambda} \rangle = \frac{1}{v_{\textrm{p},\lambda}} \sum_{j=1}^{n_\textrm{p}} V_{\textrm{p},\lambda_j}^{(j)} | b_{\textrm{p},\kappa,\lambda}^{(j)} \rangle \nonumber
\end{gather}
with dipole coupling matrix elements
\begin{gather}
    V_{\textrm{o},\kappa_i}^{(i)} = -g \langle g_{\textrm{o},\kappa_i}^{(i)} | \vec{\mu}_0^{(i)} \vec{e}  | e_{\textrm{o},\kappa_i}^{(i)} \rangle \nonumber \\
    V_{\textrm{p},\lambda_j}^{(j)} = -g \langle g_{\textrm{p},\lambda_j}^{(j)} | \vec{\mu}_0^{(j)} \vec{e} | e_{\textrm{p},\lambda_j}^{(j)} \rangle
\end{gather}
and collective coupling matrix elements
\begin{gather}
    v_{\textrm{o},\kappa} = \sqrt{\sum_{i=1}^{n_\textrm{o}}|V_{\textrm{o},\kappa_i}^{(i)}|^2} \nonumber \\
    v_{\textrm{p},\lambda} = \sqrt{\sum_{j=1}^{n_\textrm{p}}|V_{\textrm{p},\lambda_j}^{(j)}|^2}.
\label{eq:collective_coupling}
\end{gather}

\subsection{Cavity transmission spectrum}
\label{sec:transspec}

Having solved the Tavis--Cummings model for an ortho-para ensemble of $n=n_\textrm{o}+n_\textrm{p}$ molecules, we can now investigate the transmission spectrum of a Fabry--P\'erot cavity containing an  ensemble of molecules.
The transmission spectrum of Fabry--P\'erot cavities is commonly interpreted using the classical transfer-matrix approach,\cite{25McWe} which has been successfully applied to explain recent gas-phase rovibrational polariton experiments.\cite{23WrNeWe,23WrNeWe_2,24NeWe}
However, the transfer-matrix method does not provide direct access to how molecular and Fock states mix to form polaritonic states.
Therefore, in the present work, we adopt a quantum-mechanical description based on the Tavis--Cummings model to model cavity transmission spectra.
In the experiment by Weichman and co-workers, cavity-molecule strong coupling in a gas-phase sample is probed by detecting the intensity of mid-IR laser light transmitted through the cavity.\cite{23WrNeWe,23WrNeWe_2,24NeWe}
When the laser frequency is resonant with a cavity mode, the external field can efficiently drive the corresponding cavity mode. 
In the absence of molecules, the transmission spectrum consists of a series of resonances corresponding to the cavity eigenmodes.
However, if a selected cavity mode is resonantly and strongly coupled to a molecular rovibrational transition, polaritonic states emerge.
Consequently, the original cavity resonance splits into new transmission resonances whose positions reflect the energies of the polaritonic eigenstates.

Fig. \ref{fig:nmol_illustration} illustrates a Fabry--P\'erot cavity coupled to an ortho-para ensemble of $^{14}$NH$_3$ molecules.
Before describing the detailed theoretical modeling of the cavity transmission spectrum, the main steps of our approach are outlined.
First, we reiterate that molecular eigenstates are degenerate in terms of the nuclear spin ($I$ and $M_I$) and $M$ quantum numbers.
According to Section \ref{sec:tc}, polaritons can be constructed in many different ways due to these degeneracies.
Although degenerate states have the same energy, the respective light-matter coupling matrix elements are $M$-dependent (see Section \ref{sec:intmx}), which gives rise to varying Rabi splittings for polaritons with different distributions of spin and $M$ quantum numbers.
As a simplification, we now limit the discussion to polaritons for which thermodynamic probabilities, proportional to the number of possible realizations, are maximal.
In the initial state of spectral transitions, ortho and para spin isomers are assumed to occupy the ortho and para lower states
$|g_{\textrm{o},\kappa_i}^{(i)} \rangle$ and
$|g_{\textrm{p},\lambda_j}^{(j)} \rangle$,
and the cavity mode has zero photons.
From this initial state, transitions occur to final polaritonic states embedded in the singly-excited manifold.
We take the interaction between the laser field $E(t)$ and the cavity mode to be proportional to $E(t) (\hat{a}^\dagger+\hat{a})$ and apply first-order time-dependent perturbation theory to evaluate transition probabilities.

As already outlined, we consider transitions from the initial state
\begin{equation}
    | \Phi_{\kappa,\lambda} \rangle = 
        |g_{\textrm{o},\kappa_1}^{(1)},\dots,g_{\textrm{o},\kappa_{n_\textrm{o}}}^{(n_\textrm{o})},g_{\textrm{p},\lambda_1}^{(1)},\dots,g_{\textrm{p},\lambda_{n_\textrm{p}}}^{(n_\textrm{p})},0\rangle
    \label{eq:ground_state}
\end{equation}
to final states $| \Psi_{\kappa,\lambda} \rangle$ in the singly-excited manifold.
Here, ortho and para degenerate eigenstates are labeled by $\kappa_i$ and $\lambda_j$, respectively.
The computation of transition intensities requires the evaluation of transition matrix elements $\langle \Phi_{\kappa,\lambda} | \hat{a}^\dagger+\hat{a} | \Psi_{\kappa,\lambda} \rangle$.
First, transitions to bright states, that is, superpositions of bright basis functions (see Eq. \eqref{eq:bright_basis_deg})
\begin{equation}
    | \Psi_{\kappa,\lambda} \rangle = \sum_{i=0}^2 c_i |w_{i,\kappa,\lambda}\rangle
\end{equation}
are examined.
Due to the definition of bright basis functions, we get
\begin{equation}
    \langle \Phi_{\kappa,\lambda} | \hat{a}^\dagger+\hat{a} | w_{0,\kappa,\lambda} \rangle = 1
\end{equation}
and
\begin{equation}
    \langle \Phi_{\kappa,\lambda} | \hat{a}^\dagger+\hat{a} | w_{i,\kappa,\lambda} \rangle = 0
\end{equation}
for $i=1,2$, which yields
\begin{equation}
    \langle \Phi_{\kappa,\lambda} | \hat{a}^\dagger+\hat{a} | \Psi_{\kappa,\lambda} \rangle = c_0.
\label{eq:tmx}
\end{equation}
In other words, the transition matrix element of Eq. \eqref{eq:tmx} measures the photonic component of the final state $| \Psi_{\kappa,\lambda} \rangle$.
Since dark states have zero photonic component, transition matrix elements between $| \Phi_{\kappa,\lambda} \rangle$ and dark states are identically zero.
The transmission intensity
\begin{equation}
    I \sim | \langle \Phi_{\kappa,\lambda} | \hat{a}^\dagger+\hat{a} | \Psi_{\kappa,\lambda} \rangle |^2
\end{equation}
is taken to be proportional to the absolute square of the transition matrix element.

As explained in the previous sections, rovibrational eigenstates of molecules are typically degenerate.
For the eigenstates of $^{14}$NH$_3$ examined in this work, degeneracies occur in terms of the nuclear spin quantum numbers $I$ and $M_I$ as well as the quantum number $M$.
Here, a canonical ensemble of ortho and para spin isomers is considered at temperature $T$.
The thermodynamic probability $W_{\kappa,\lambda}$ of the initial state $| \Phi_{\kappa,\lambda} \rangle$ defined in Eq. \eqref{eq:ground_state} is proportional to
\begin{equation}
    W_{\kappa,\lambda} \sim 
    \frac{n!}{\prod_{\kappa=1}^{d_\textrm{o}^g} n_{\textrm{o},\kappa}! \prod_{\lambda=1}^{d_\textrm{p}^g} n_{\textrm{p},\lambda}!}
    \exp[-\beta (n_\textrm{o} E_\textrm{o}^{(g)} + n_\textrm{p} E_\textrm{p}^{(g)})]
    \label{eq:tdprob}
\end{equation}
where 
$\kappa = 1, \dots, d_\textrm{o}^g$, 
$\lambda = 1, \dots, d_\textrm{p}^g$,
the number of ortho and para molecules occupying states
$|g_{\textrm{o},\kappa} \rangle$ and $|g_{\textrm{p},\lambda} \rangle$
equals $n_{\textrm{o},\kappa}$ and $n_{\textrm{p},\lambda}$, respectively, 
$n_\textrm{o} = \sum_{\kappa=1}^{d_\textrm{o}^g} n_{\textrm{o},\kappa}$,
$n_\textrm{p} = \sum_{\lambda=1}^{d_\textrm{p}^g} n_{\textrm{p},\lambda}$
and $\beta = 1/(k_\textrm{B} T)$.
Using Stirling's approximation for a large number of molecules $n$, one can show that $W_{\kappa,\lambda}$ is maximal under the constraint
$\sum_{\kappa=1}^{d_\textrm{o}^g} n_{\textrm{o},\kappa} + \sum_{\lambda=1}^{d_\textrm{p}^g} n_{\textrm{p},\lambda} = n$ if
\begin{gather}
    n_{\textrm{o},\kappa'} = \frac{n}{d_\textrm{o}^g + d_\textrm{p}^g \exp(-\beta(E_\textrm{p}^{(g)}-E_\textrm{o}^{(g)}))} \nonumber \\
    n_{\textrm{p},\lambda'} = \frac{n}{d_\textrm{o}^g \exp(-\beta(E_\textrm{o}^{(g)}-E_\textrm{p}^{(g)})) + d_\textrm{p}^g}
\label{eq:wmax1}
\end{gather}
and
\begin{gather}
    n_\textrm{o}' = \sum_{\kappa'=1}^{d_\textrm{o}^g} n_{\textrm{o},\kappa'} =\frac{n d_\textrm{o}^g}{d_\textrm{o}^g + d_\textrm{p}^g \exp(-\beta(E_\textrm{p}^{(g)}-E_\textrm{o}^{(g)}))} \nonumber \\
    n_\textrm{p}' = \sum_{\lambda'=1}^{d_\textrm{p}^g} n_{\textrm{p},\lambda'} = \frac{n d_\textrm{p}^g}{d_\textrm{o}^g \exp(-\beta(E_\textrm{o}^{(g)}-E_\textrm{p}^{(g)})) + d_\textrm{p}^g}.
\label{eq:wmax2}
\end{gather}
Here, the primes refer to quantities that maximize $W_{\kappa,\lambda}$.

In the following, we limit the discussion to an initial state
\begin{equation}
    | \Phi_{\kappa',\lambda'} \rangle = 
        |g_{\textrm{o},\kappa_1'}^{(1)},\dots,g_{\textrm{o},\kappa_{n_\textrm{o}'}'}^{(n_\textrm{o}')},g_{\textrm{p},\lambda_1'}^{(1)},\dots,g_{\textrm{p},\lambda_{n_\textrm{p}'}'}^{(n_\textrm{p}')},0\rangle
    \label{eq:ground_state_max}
\end{equation}
that satisfies Eq. \eqref{eq:wmax1} and Eq. \eqref{eq:wmax2}.
In other words, $W_{\kappa',\lambda'}$ is maximal.
Due to the orthogonality of molecular eigenstates, the operator $\hat{a}^\dagger+\hat{a}$ can induce transitions to final states $| \Psi_{\kappa',\lambda'} \rangle$ that have nonzero overlap with the state
\[
    | b_{\kappa',\lambda'} \rangle =
    |g_{\textrm{o},\kappa_1'}^{(1)},\dots,g_{\textrm{o},\kappa_{n_\textrm{o}'}'}^{(n_\textrm{o}')},g_{\textrm{p},\lambda_1'}^{(1)},\dots,g_{\textrm{p},\lambda_{n_\textrm{p}'}'}^{(n_\textrm{p}')},1\rangle.
\]
According to Section \ref{sec:model}, dipole moment matrix elements are diagonal in terms of $M$, $I$ and $M_I$.
Keeping in mind that labels $\kappa_i$ and $\lambda_j$ refer to $M$, $I$ and $M_I$ in our case,
one can see that $| b_{\kappa',\lambda'} \rangle$ can be coupled only to states
\[
    |g_{\textrm{o},\kappa_1'}^{(1)},\dots,e_{\textrm{o},\kappa_i'}^{(i)},\dots,g_{\textrm{o},\kappa_{n_\textrm{o}'}'}^{(n_\textrm{o}')},g_{\textrm{p},\lambda_1'}^{(1)},\dots,g_{\textrm{p},\lambda_{n_\textrm{p}'}'}^{(n_\textrm{p}')},0\rangle
\]
and
\[
   |g_{\textrm{o},\kappa_1'}^{(1)},\dots,g_{\textrm{o},\kappa'_{n_\textrm{o}'}}^{(n_\textrm{o}')},g_{\textrm{p},\lambda_1'}^{(1)},\dots,e_{\textrm{p},\lambda_j'}^{(j)},\dots,g_{\textrm{p},\lambda_{n_\textrm{p}'}}^{(n_\textrm{p}')},0\rangle
\]
by the dipole interaction.
The relations established in this part will be used to determine collective coupling matrix elements $v_{\textrm{o},\kappa}$ and $v_{\textrm{p},\lambda}$ defined in Eq. \eqref{eq:collective_coupling}.
We stress that since dipole moment matrix elements depend on the quantum number $M$ (see Section \ref{sec:model}), $v_{\textrm{o},\kappa}$ and $v_{\textrm{p},\lambda}$ are sensitive to the choice of quantum numbers $\kappa$ and $\lambda$ that have been introduced to label degenerate eigenstates.
In this paper, $v_{\textrm{o},\kappa}$ and $v_{\textrm{p},\lambda}$ are consistently evaluated using Eq. \eqref{eq:wmax1} and Eq. \eqref{eq:wmax2}.
The investigation of effects related to polaritonic branches other than the one maximizing $W_{\kappa,\lambda}$ is left for future work.
As a next step, bright polaritonic states can be determined by diagonalizing the bright Hamiltonian of Eq. \eqref{eq:mx_bright}.
With the energies and photonic components of the bright polaritonic states at hand, one can simulate the cavity transmission spectrum for the ortho-para ensemble as described in this section.

\section{Pauli principle and symmetry}
\label{sec:symm}

We choose the $^{14}$NH$_3$ molecule for our investigations.
$^{14}$NH$_3$ consists of three identical spin-1/2 protons which can be coupled to a total nuclear spin of $I=3/2$ or $I=1/2$, resulting in ortho ($I=3/2$) and para ($I=1/2$) spin isomers.
In addition, the nuclear spin of $^{14}$N equals $I=1$.
The molecular symmetry (MS) group of $^{14}$NH$_3$ is $D_{3\textrm{h}}(\textrm{M})$ \cite{06BuJe} which has the carrier set $\{E,(12),(13),(23),(123),(132)\} \otimes \{E,E^*\}$.
Here, $(ij)$ denotes the permutation of protons $i$ and $j$ (transposition), $(123)$ and $(132)$ refer to the two possible cyclic permutations of the three protons and $^*$ is the space inversion operation ($x \rightarrow -x$, $y \rightarrow -y$ and $z \rightarrow -z$).
The rovibrational Hamiltonian of $^{14}$NH$_3$ commutes with the elements of $D_{3\textrm{h}}(\textrm{M})$. 
The character table of $D_{3\textrm{h}}(\textrm{M})$ is given in Table SI of the Supporting Information.

According to the Pauli principle, rovibrational eigenstates of $^{14}$NH$_3$ must change sign under the exchange of two arbitrary protons (fermions).
Spatial eigenstates need to be combined with
nuclear spin functions and then properly antisymmetrized to satisfy the Pauli principle.
By coupling three 1/2-spins, eight spin states can be formed, a quadruplet $|I,M_I\rangle$ with $I=3/2$ and $M_I = \pm 3/2, \pm 1/2$, and 
two distinct doublets $|I,M_I^{(i)}\rangle$ with $I=1/2$, $M_I = \pm 1/2$ and $i=1,2$.
These spin functions can be constructed using Clebsch--Gordan coefficients and they are listed in Table SII of the Supporting Information
together with nuclear spin quantum numbers ($I$ and $M_I=-I,\dots,I$).
Group-theoretical considerations reveal that the eight spin functions span the representation $\Gamma_\textrm{spin} = 4 \textrm{A}_1' \oplus 2 \textrm{E}'$.\cite{06BuJe}
More precisely, spin functions $|3/2,M_I\rangle$ with a given value of $M_I = \pm 3/2, \pm 1/2$ span irreducible representation (irrep) $\textrm{A}_1'$, while doublets $|1/2,M_I^{(i)}\rangle$ with $i=1,2$ span irrep $\textrm{E}'$ for $M_I = 1/2$ and $M_I = -1/2$.

Pauli-allowed eigenstates of $^{14}$NH$_3$ transform according to irreps $\textrm{A}_2'$ or $\textrm{A}_2''$ of $D_{3\textrm{h}}(\textrm{M})$.\cite{06BuJe}
Combining the symmetries $\Gamma_\textrm{spatial}$ and $\Gamma_\textrm{spin}$ of spatial and spin eigenfunctions, it is easy to see that the representation $\Gamma_\textrm{tot} = \Gamma_\textrm{spatial} \otimes \Gamma_\textrm{spin}$ must contain irreps $\textrm{A}_2'$ or $\textrm{A}_2''$.
If $\Gamma_\textrm{spatial} = \textrm{A}_1'/\textrm{A}_1''$, no spin function is compatible with the spatial part.
As a consequence, eigenstates having $\Gamma_\textrm{spatial} = \textrm{A}_1'/\textrm{A}_1''$
are Pauli forbidden. If $\Gamma_\textrm{spatial} = \textrm{A}_2'/\textrm{A}_2''$, spin functions of $\textrm{A}_1'$ symmetry
give $\Gamma_\textrm{tot} = \textrm{A}_2'/\textrm{A}_2''$, which means that each eigenstate with $\Gamma_\textrm{spatial} = \textrm{A}_2'/\textrm{A}_2''$
is compatible with four spin functions of $I=3/2$.
Finally, if $\Gamma_\textrm{spatial} = \textrm{E}'/\textrm{E}''$, spin functions of $\textrm{E}'$ symmetry ($I =1/2$) are appropriate
since 
$\Gamma_\textrm{tot} = \textrm{E}' \otimes \textrm{E}' = \textrm{A}_1' \oplus \textrm{A}_2' \oplus \textrm{E}'$ and
$\Gamma_\textrm{tot} = \textrm{E}' \otimes \textrm{E}'' = \textrm{A}_1'' \oplus \textrm{A}_2'' \oplus \textrm{E}''$
contain irreps $\textrm{A}_2'$ and $\textrm{A}_2''$, respectively.

If multiple molecules are considered, further (anti)symmetrization of the wave function is required.
In case of $^{14}$NH$_3$ molecules, the exchange of two molecules involves the exchange of three pairs of fermions (protons) and one pair of bosons ($^{14}\textrm{N}$ nuclei).
Therefore, $^{14}$NH$_3$ behaves as a composite fermion and Pauli-allowed states must be antisymmetric under the exchange of any two molecules.
This property naturally gives rise to exchange effects which become irrelevant if molecules are far apart and molecular wave functions have virtually zero overlaps.\cite{17SaNa}
In what follows, permutational symmetries of identical nuclei are enforced consistently within each molecule.
However, we assume spatially well-separated molecules and neglect effects associated with the exchange of identical molecules.
Thus, molecular ensembles are described using product-state wave functions in our model.
Note that if one considers an ortho and a para spin isomer, exchange terms are exactly zero due to the orthogonality of ortho and para spin states, and the two spin isomers behave effectively distinguishably.

\section{Computational details}
\label{sec:model}

\subsection{Molecular model}

$^{14}$NH$_3$, a prototypical system for quantum tunneling, has two symmetry-equivalent pyramidal equilibrium structures which  are connected by a planar transition-state structure along the inversion path (umbrella motion).
The interconversion of the two pyramidal structures can be described by a symmetric double-well potential.
Our full-dimensional computational approach includes all rotational and vibrational degrees of freedom.
Numerically exact rovibrational energy levels and eigenstates have been obtained by variational nuclear-motion computations using the GENIUSH program,\cite{09MaCzCs,11FaMaCs} a full-dimensional potential surface,\cite{11YuBaTe} Radau polyspherical coordinates \cite{80Smith} and a direct-product discrete variable representation (DVR) \cite{00LiCa} vibrational basis.
Our rovibrational energy levels show excellent agreement with the experimental values \cite{25SmYuTe} for the eigenstates considered in this work.
GENIUSH energy levels are used in all numerical computations in this work.

We focus on rovibrational transitions associated with the inversion ($\nu_2$) vibration which gives a high-intensity absorption band in the IR spectrum, facilitating strong coupling to the cavity mode.
Keeping in mind that tunneling in a symmetric double-well potential leads to the splitting of energy levels, and the eigenstates are either symmetric or antisymmetric with respect to inversion,
rovibrational eigenstates of $^{14}$NH$_3$ are labeled $|v^pJKM\rangle$.
Here, $v=0,1,2,\dots$ is the inversion quantum number, $p=\pm$ denotes the vibrational parity, while 
$J = 0, 1, 2, \dots$ (rotational quantum number),
$K = -J,\dots,0,\dots,J$ (projection along the body-fixed (BF) $z$ axis) and 
$M = -J,\dots,0,\dots,J$ (projection along the space-fixed (SF) $z$ axis) refer to symmetric top quantum numbers.\cite{06BuJe}
At the same time, we assume that vibrational modes other than the inversion mode are not excited (the respective vibrational quantum numbers equal $v_i=0$).
Therefore, the molecular eigenstate basis used in numerical computations is restricted to inversion-rotation eigenstates $|v^pJKM\rangle$.
Hyperfine interactions are neglected, which implies that Hamiltonian matrix elements between ortho and para states vanish.
It is worth noting that eigenstates $|v^pJKM\rangle$ differing only in the value of $M$ are degenerate.

The evaluation of light-matter interaction matrix elements requires quantum-chemical computations in the electronic ground state of $^{14}$NH$_3$.
First, the minimum-energy inversion path has been mapped by changing the inversion angle and optimizing the remaining five internal coordinates along the inversion path.
These computations employed the Gaussian 16 program,\cite{16FrTrSc} and resulted in dipole moments and polarizabilities along the inversion path at the CCSD/aug-cc-pVTZ level of theory.
In addition, DSE computations were carried out along the inversion path using an in-house program \cite{25FaHaHo} at the CCSD/aug-cc-pVTZ level of theory. 
The necessary integrals as well as CCSD one-electron and two-electron reduced density matrices were evaluated with the PySCF program.\cite{18SuBeBl,20SuZhBa}
We stress that our DSE computations result in an accurate DSE surface without making approximations used by others.\cite{24YuBo,24BoScKo}
The inversion path maintains $C_{3\textrm{v}}$ point group symmetry and the BF frame corresponds to the principal axis frame.
The principal rotation axis is chosen as the BF $z$ axis, the N atom is placed on the BF $z$ axis and the first H atom is located in the BF $xz$ plane (with $x>0$).
Due to symmetry, the dipole moment vector is aligned with the BF $z$ axis
($\mu^\textrm{BF}_x = \mu^\textrm{BF}_y = 0$),
while the polarizability and DSE tensors are diagonal with
$\alpha^\textrm{BF}_{xx} = \alpha^\textrm{BF}_{yy} \ne \alpha^\textrm{BF}_{zz}$
and
$d^\textrm{BF}_{xx} = d^\textrm{BF}_{yy} \ne d^\textrm{BF}_{zz}$
in our model.
Relevant electronic structure data are available in Table SIII of the Supporting Information.

\subsection{Spin degeneracy factors}

Using group theory, one can determine which irreps of $D_{3\textrm{h}}(\textrm{M})$ are spanned by the inversion-rotation eigenfunctions $|v^pJKM\rangle$. 
We refer to Chapter 15 of Ref. \citenum{06BuJe} for detailed symmetry considerations and summarize the symmetry properties of $|v^pJKM\rangle$ in Table \ref{tbl:symmprop}.
For $K > 0$, eigenfunctions $|v^pJKM\rangle$ and $|v^pJ\!-\!\!KM\rangle$ 
span two-dimensional irreps for $K ~ \textrm{mod} ~ 3 \ne 0 $ and the direct sum of two one-dimensional irreps for $K ~ \textrm{mod} ~ 3 = 0 $.
Pauli-allowed eigenstates of $^{14}$NH$_3$ are characterized in Table \ref{tbl:pauli} where possible values of vibrational, rotational and spin quantum numbers are also specified.
These rules can be derived by using the symmetry considerations presented in Section \ref{sec:symm}.
For $K=0$, combinations of even $J$ values with vibrational parity $p=+$ and odd $J$ values with $p=-$ are Pauli-forbidden
($\Gamma_\textrm{spatial} = \textrm{A}_1'/\textrm{A}_1''$).
Moreover, the ortho ($I=3/2$) and para ($I=1/2$) spin isomers correspond to $K ~ \textrm{mod} ~ 3 = 0$
(only $\Gamma_\textrm{spatial} = \textrm{A}_2'/\textrm{A}_2''$)
and $K ~ \textrm{mod} ~ 3 \ne 0$ 
($\Gamma_\textrm{spatial} = \textrm{E}'/\textrm{E}''$),
respectively.
Using the transformation properties of spatial and spin eigenfunctions under symmetry elements of $D_{3\textrm{h}}(\textrm{M})$, Pauli-allowed eigenstates can be constructed by group-theoretical projectors.

With these rules at hand, spin degeneracy factors can be readily determined for $^{14}$NH$_3$.
For Pauli-allowed rovibrational eigenstates with $K=0$, each spatial eigenstate is compatible with four $I=3/2$ nuclear spin functions, yielding a spin degeneracy factor of four.
If $K>0$ and $K ~ \textrm{mod} ~ 3 = 0$, eigenstate pairs with quantum numbers $K$ and $-K$ are not strictly degenerate and spatial eigenstates of $\textrm{A}_2'$ or $\textrm{A}_2''$ symmetry are compatible with four $I=3/2$ spin functions.
Finally, for $K ~ \textrm{mod} ~ 3 \ne 0$, eigenstate pairs with $K$ and $-K$ are degenerate and two degenerate spatial eigenstates combined with two $I=1/2$ spin functions give a single Pauli-allowed eigenstate.
Since there are two $I=1/2$ spin function pairs, unique energy levels (that is, without counting spatial degeneracies) carry a spin degeneracy factor of two.

\begin{table}
  \caption{\label{tbl:symmprop}
  Classification of inversion-rotation eigenfunctions $|v^pJKM\rangle$ of the $^{14}$NH$_3$ molecule according to irreducible representations $\Gamma_\textrm{spatial}$ of the $D_{3\textrm{h}}(\textrm{M})$ molecular symmetry group.
  Quantum numbers relevant for symmetry classification are the rotational quantum number ($J$), the angular momentum projection quantum number along the body-fixed $z$ axis ($K$) and the vibrational parity ($p$).
  $K \ge 0$ and $n$ is a nonnegative integer.}
  \begin{tabular}{c|c|c|c|c|c|c}
          $p$ & $K=0$, $J$ even & $K=0$, $J$ odd & $K=6n+3$ & $K=6n+6$ & $K=6n \pm 2$ & $K=6n \pm 1$ \\
        \hline
    $+$ &  $\textrm{A}_1'$ & $\textrm{A}_2'$ & $\textrm{A}_1'' \oplus \textrm{A}_2''$ & $\textrm{A}_1' \oplus \textrm{A}_2'$ & 
    $\textrm{E}'$ & \ $\textrm{E}''$ \\
    $-$ &  $\textrm{A}_2''$ & $\textrm{A}_1''$ & $\textrm{A}_2' \oplus \textrm{A}_1'$ & $\textrm{A}_2'' \oplus \textrm{A}_1''$ & 
    $\textrm{E}''$ & \ $\textrm{E}'$
  \end{tabular}
\end{table}

\begin{table}
  \caption{\label{tbl:pauli}
  Pauli-allowed rovibrational eigenstates of the $^{14}$NH$_3$ molecule labeled with symmetric top quantum numbers ($J$, $K$ and $M$), vibrational quantum number of the inversion mode ($v^p$, where $p=\pm$ is the vibrational parity) and nuclear spin quantum numbers ($I$ and $M_I=-I,\dots,I$).
  $I=3/2$ and $I=1/2$ correspond to the ortho and para spin isomers of $^{14}$NH$_3$, respectively.
  The definition of spin eigenfunctions with $I=3/2$ and $I=1/2$ can be found in Table SII of the Supporting Information.
  $J=0,1,2,\dots$, $v^p=0^p,1^p,2^p,\dots$ and $K\ge0$ in the table.}
  \begin{tabular}{c|c|c|c|c|c|c}
      $J$ & $K$  & $M$ & $v^p$ & $I$ & $M_I$ & spin isomer \\
      \hline
      even & $K=0$ & $-J,\dots,J$ & $v^-$ & $3/2$ & $\pm 3/2, \pm 1/2$ & ortho   \\
      odd & $K=0$ & $-J,\dots,J$ & $v^+$ & $3/2$ & $\pm 3/2, \pm 1/2$ & ortho   \\
      even/odd & $K ~ \textrm{mod} ~ 3 = 0$, $K>0$ & $-J,\dots,J$ & $v^\pm$ & $3/2$ & $\pm 3/2, \pm 1/2$ & ortho \\
      even/odd & $K ~ \textrm{mod} ~ 3 \ne 0$ & $-J,\dots,J$ & $v^\pm$ & $1/2$ & $ \pm 1/2$ &  para
  \end{tabular}
\end{table}

\subsection{Light-matter interaction matrix elements}
\label{sec:intmx}

Light-matter interaction (dipole moment, polarizability and DSE) matrix elements are evaluated using an effective inversion-rotation model based on a one-dimensional inversion potential.\cite{97LeUt}
In what follows, the cavity polarization vector is chosen as $\mathbf{e}=(0,0,1)$, referenced in the SF frame.
Rovibrational eigenstates $|v^pJKM\rangle$ are approximated as products of one-dimensional inversion eigenfunctions $|v^p\rangle$ and symmetric-top eigenfunctions $|JKM\rangle$.
The one-dimensional inversion problem is defined by the Hamiltonian
\begin{equation}
    \hat{H}_\textrm{inv} = -\frac{\hbar^2}{2m} \frac{\textrm{d}^2}{\textrm{d}^2x} + V(x)
\end{equation}
where $x$ is the distance from the planar transition-state structure along the arc moved by the H atoms, the effective tunneling mass equals $m = 2.56217 ~ \textrm{u}$ and $V(x)$ is an empirical double-well potential designed for the accurate description of inversion tunneling in $^{14}$NH$_3$.\cite{97LeUt}
The validity of this approximation was verified by comparison to full-dimensional rovibrational results obtained with GENIUSH.

Next, matrix elements of the $z$ component of the SF dipole moment are expressed in the approximate $|v^pJKM\rangle$ eigenbasis, which reads
\begin{gather}
    \langle v^pJKM| \mu^\textrm{SF}_z |v'^{p'}J'K'M'\rangle = 
    (-1)^{M-K} \sqrt{(2J+1)(2J'+1)} \nonumber \\
    \langle v^{p} | \mu^\textrm{BF}_z | v'^{p'} \rangle
    \begin{pmatrix}
        J & 1 & J'\\
       -M & 0 & M'
    \end{pmatrix} 
    \begin{pmatrix}
        J & 1 & J'\\
       -K & 0 & K'
    \end{pmatrix}
    \label{eq:dipmx}
\end{gather}
where Wigner's 3-j symbol \cite{17SaNa} is denoted by
\[
\begin{pmatrix}
    j_1 & j_2 & j_3\\
    m_1 & m_2 & m_3
\end{pmatrix}
\]
(see the Supporting Information for more information).
Since the DSE and polarizability tensors are symmetric tensors of rank two, we introduce a common notation $T$ for them.
As shown in the Supporting Information, matrix elements of the DSE and polarizability tensors $T^\textrm{SF}_{zz}$ in the approximate $|v^{p}JKM\rangle$ eigenbasis read
\begin{gather}
    \langle v^{p}JKM| T^\textrm{SF}_{zz} |v'^{p'}J'K'M'\rangle = 
    \langle v^{p} | \frac{2T^\textrm{BF}_{xx}+T^\textrm{BF}_{zz}}{3} | v'^{p'} \rangle \delta_{JJ'} \delta_{KK'} \delta_{MM'} + (-1)^{M-K} \nonumber \\
    \sqrt{(2J+1)(2J'+1)} 
    \langle v^{p} | \frac{2(T^\textrm{BF}_{zz}-T^\textrm{BF}_{xx})}{3} | v'^{p'} \rangle
    \begin{pmatrix}
        J & 2 & J'\\
       -M & 0 & M'
    \end{pmatrix}
    \begin{pmatrix}
        J & 2 & J'\\
       -K & 0 & K'
    \end{pmatrix}.
    \label{eq:dsemx}
\end{gather}
The 3-j symbols in Eq. \eqref{eq:dipmx} and Eq. \eqref{eq:dsemx} imply that selection rules $M=M'$ and $K=K'$ hold for $\mathbf{e} = (0,0,1)$.
The necessary BF light-matter interaction matrix elements
$\langle v^p | \mu^\textrm{BF}_z |v'^{p'}\rangle$,
$\langle v^p | \alpha^\textrm{BF}_{xx} |v'^{p'}\rangle$,
$\langle v^p | \alpha^\textrm{BF}_{zz} |v'^{p'}\rangle$,
$\langle v^p | d^\textrm{BF}_{xx} |v'^{p'}\rangle$ 
and
$\langle v^p | d^\textrm{BF}_{zz} |v'^{p'}\rangle$
can be readily evaluated using the one-dimensional inversion eigenfunctions of $\hat{H}_\textrm{inv}$ represented in a one-dimensional DVR basis.
The conservation of $M$ is exact, while $K$ is conserved only in our approximate model.
Test computations using numerically exact GENIUSH rovibrational eigenstates and the full-dimensional dipole moment surface of Ref. \citenum{03MaQuTh} confirm that dipole matrix elements violating the conservation of $K$ have substantially smaller absolute values than dipole matrix elements with $\Delta K=0$.
These tests serve as a validation of the approximate model used to treat light-matter interaction terms.

\subsection{Time-dependent quantum dynamics}
\label{sec:tdqm}

The time evolution of the coupled cavity-molecule system is described by the Lindblad master equation \cite{20Manzano}
\begin{equation}
    \frac{\partial \hat{\rho}}{\partial t} = -\frac{\textrm{i}}{\hbar} [\hat{H},\hat{\rho}] +
    \gamma_\textrm{c} \hat{a} \hat{\rho} \hat{a}^\dagger - \frac{\gamma_\textrm{c}}{2} (\hat{\rho} \hat{N} + \hat{N} \hat{\rho} )
\label{eq:Lindblad}
\end{equation}
where $\hat{\rho}$ is the density operator, $\hat{N} = \hat{a}^\dagger \hat{a}$ gives the number operator
and $\gamma_\textrm{c}$ is the cavity decay rate equivalent to a 
lifetime of $\tau = 1/\gamma_\textrm{c}$.
This way, finite lifetimes of cavity excitations and incoherent decay effects are taken into account.
At the same time, lifetimes of excited molecular eigenstates are assumed to be infinite, which is plausible on the $50 ~ \textrm{ps}$ timescale used in Section \ref{sec:2molres}.
The Lindblad equation is integrated numerically in the eigenstate basis of $\hat{H}$ using the interaction picture and a fourth-order Runge--Kutta scheme.
To this end, the cavity-molecule Hamiltonian is first diagonalized in the direct product basis of Pauli-allowed molecular eigenstates and Fock states of the cavity mode $|n\rangle$ (with $n=0,1,2,\dots$).
In all numerical computations, Fock states with $n=0,\dots,3$ were included in the basis.
For the two-molecule case investigated in Section \ref{sec:2molres}, the basis is constructed as the direct product of Pauli-allowed eigenstates of the first and second molecules, and $|n\rangle$.
The diagonalization of the cavity-molecule Hamiltonian results in energy levels and eigenstates (polaritonic states) of the coupled cavity-molecule system.
As already discussed in Section \ref{sec:intmx},  both the $K$ and $M$ quantum numbers are conserved for cavity polarization $\mathbf{e} = (0,0,1)$ used in this work.
Therefore, non-interacting blocks of the cavity-molecule Hamiltonian can be diagonalized  separately for different values of $K$ and $M$.

\section{Two-molecule results}
\label{sec:2molres}

First, we investigate the case of two $^{14}$NH$_3$ molecules coupled to a cavity mode and prepared in appropriately selected initial quantum states.
All results presented in this section are consistent with the Hamiltonian of Eq. \eqref{eq:Hcm_gs_mult} including the DSE and polarizability terms.
We assume a plasmonic cavity setup and employ cavity parameters that correspond to an effective mode volume of $V_\textrm{eff} \approx 400 ~ \textrm{nm}^3$.
According to Ref. \citenum{22MoSeOc}, $V_\textrm{eff}$ values less than $100 ~ \textrm{nm}^3$ should be feasible in IR plasmonic cavities.
Quality factors listed in Table 1 of Ref. \citenum{22MoSeOc} are on the order of $Q\approx10-20$.
These values fall short of $Q \approx 900$ which was adopted by our model to access VSC for two molecules.
In addition, we assume a spatially uniform cavity field, which implies that the same coupling strength parameter is used for both molecules.
We stress that molecular rotations are included in our model and the two molecules do not experience the same coupling due to different transition dipole matrix elements.
By artificially increasing $Q$ in our numerical computations, we can observe several periods of Rabi oscillation before the system relaxes
and we can identify striking effects related to the Pauli principle in polaritonic chemistry.

\begin{figure}[ht]
\includegraphics[scale=0.5]{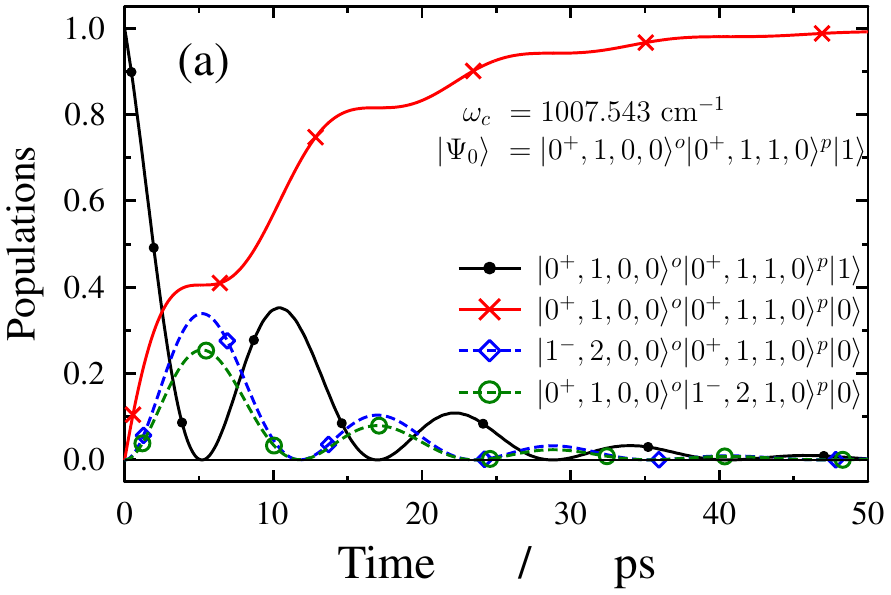}
\includegraphics[scale=0.5]{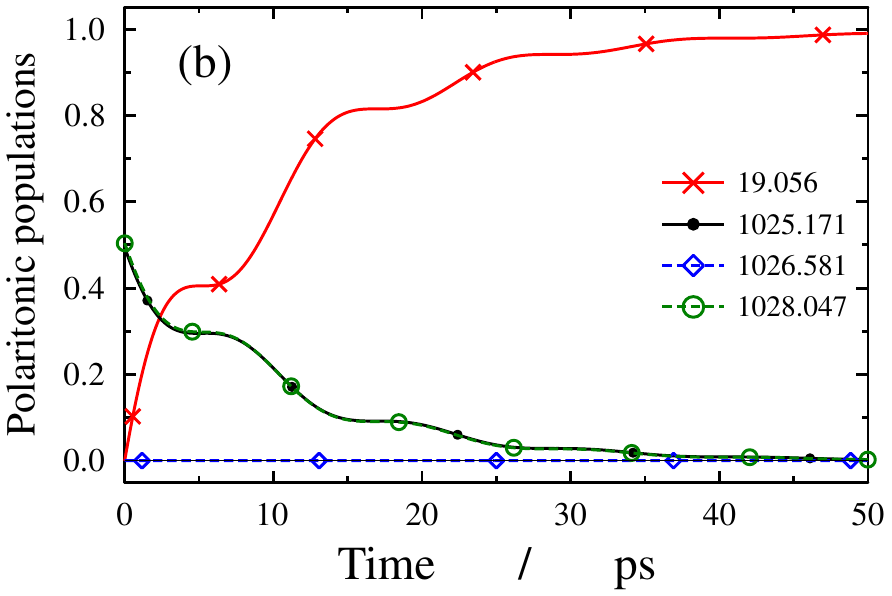}
\includegraphics[scale=0.5]{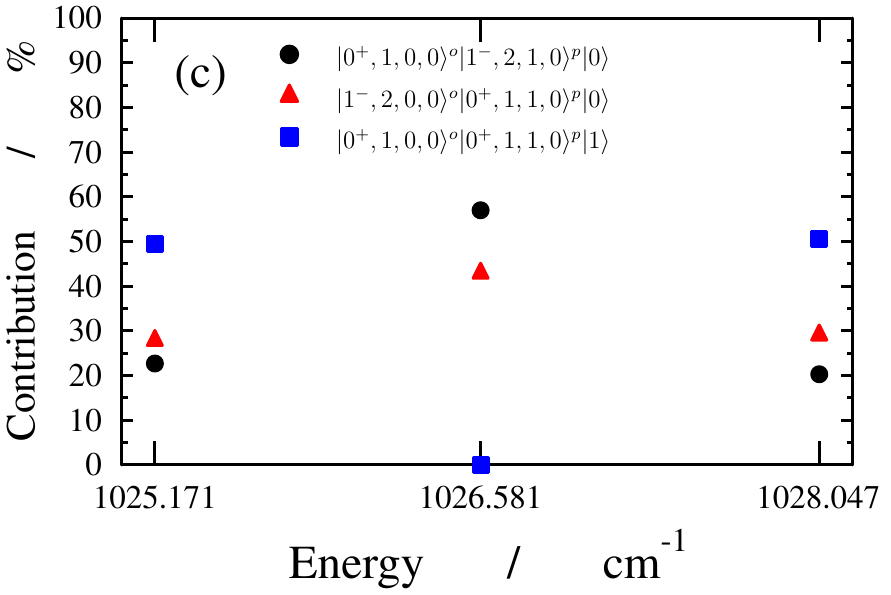}
\caption{\label{fig:2mol_1}
Time-dependent populations of selected states
$|v_1^{p_1}J_1K_1M_1\rangle^{\textrm{i}_1} |v_2^{p_2}J_2K_2M_2\rangle^{\textrm{i}_2} |n\rangle$
(panel (a)) and polaritonic states (panel (b)) as well as decomposition of polaritonic states in the $|v_1^{p_1}J_1K_1M_1\rangle^{\textrm{i}_1} |v_2^{p_2}J_2K_2M_2\rangle^{\textrm{i}_2} |n\rangle$ basis (panel (c)).
In all panels, one ortho and one para spin isomer of $^{14}$NH$_3$ are coupled to the cavity.
States are labeled with vibrational ($v^p$, where $v$ is the inversion quantum number and $p$ is the vibrational parity) and symmetric top ($J$, $K$ and $M$) quantum numbers, while $|n\rangle$ denotes Fock states of the cavity mode and $\textrm{i} = \textrm{o}, \textrm{p}$ refers to ortho and para spin isomers, respectively.
Energies of the selected polaritonic states are given in the legend of panel (b) and on the $x$ axis of panel (c) in units of $\textrm{cm}^{-1}$.
Cavity parameters are chosen as $\omega_\textrm{c}=1007.543 ~ \textrm{cm}^{-1}$, $g=10^{-4} ~ \textrm{au}$ and $\tau = 5 ~ \textrm{ps}$.
In this figure, one can observe the formation of collective ortho-para polaritons.}
\end{figure}

\begin{figure}[ht]
\includegraphics[scale=0.5]{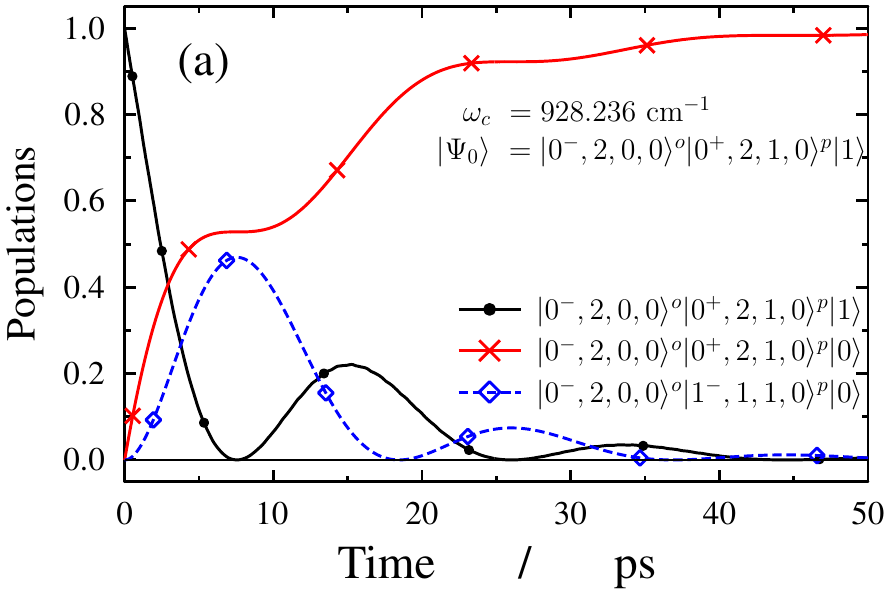}
\includegraphics[scale=0.5]{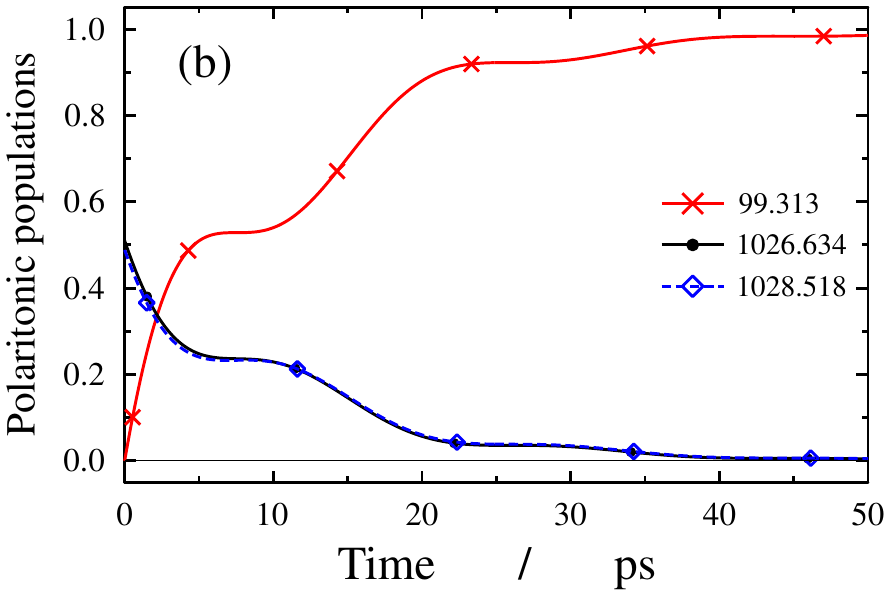}
\includegraphics[scale=0.5]{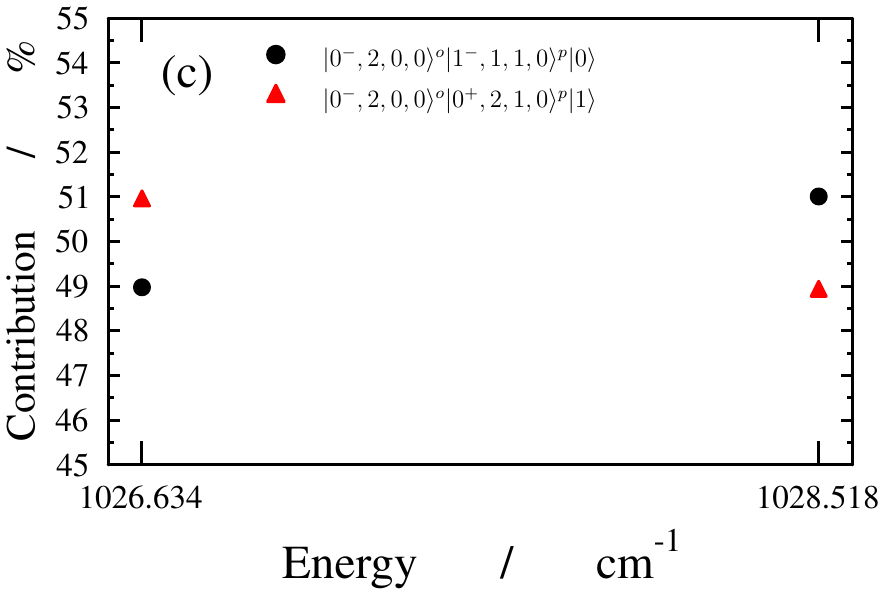}
\caption{\label{fig:2mol_2}
Time-dependent populations of selected states
$|v_1^{p_1}J_1K_1M_1\rangle^{\textrm{i}_1} |v_2^{p_2}J_2K_2M_2\rangle^{\textrm{i}_2} |n\rangle$
(panel (a)) and polaritonic states (panel (b)) as well as decomposition of polaritonic states in the $|v_1^{p_1}J_1K_1M_1\rangle^{\textrm{i}_1} |v_2^{p_2}J_2K_2M_2\rangle^{\textrm{i}_2} |n\rangle$ basis (panel (c)).
In all panels, one ortho and one para spin isomer of $^{14}$NH$_3$ are coupled to the cavity.
The notations are the same as in Fig. \ref{fig:2mol_1}.
Cavity parameters are chosen as $\omega_\textrm{c}=928.236 ~ \textrm{cm}^{-1}$, $g=10^{-4} ~ \textrm{au}$ and $\tau = 5 ~ \textrm{ps}$.
In this case, only the para isomer can couple to the cavity and the ortho isomer acts as a spectator.}
\end{figure}

\begin{figure}[ht]
\includegraphics[scale=0.5]{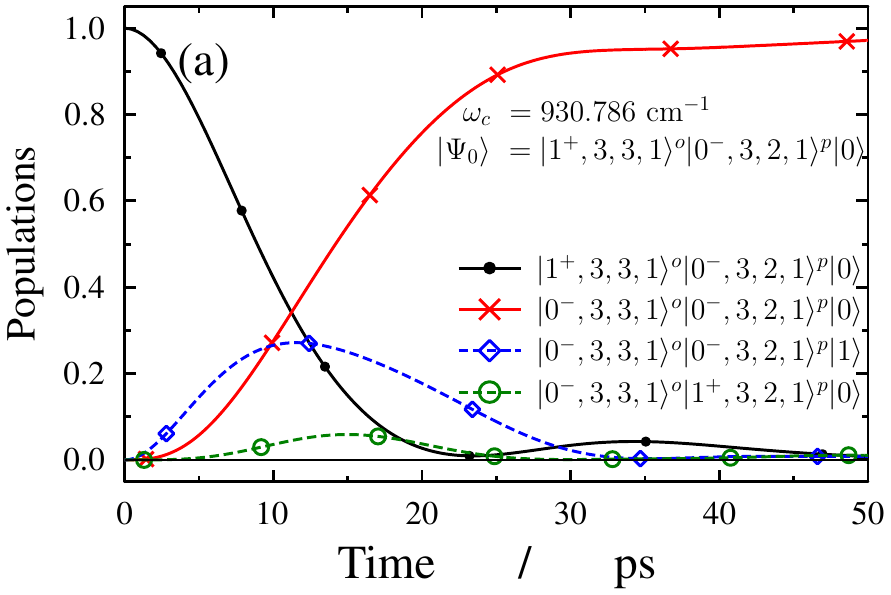}
\includegraphics[scale=0.5]{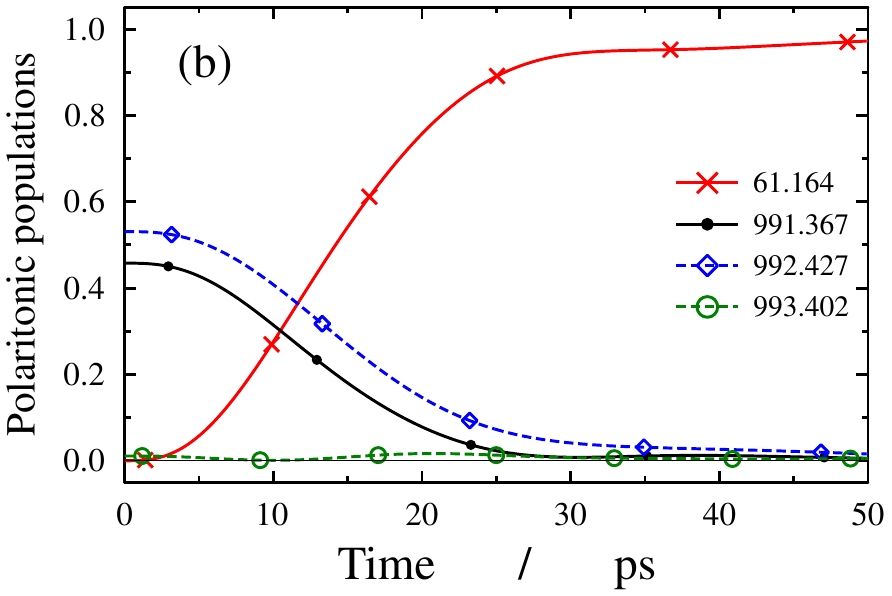}
\includegraphics[scale=0.5]{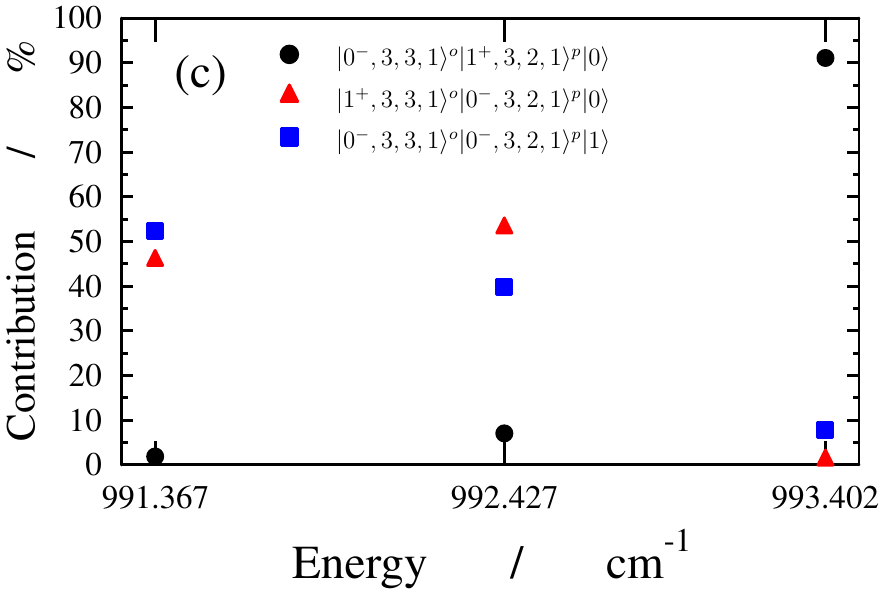}
\caption{\label{fig:2mol_3}
Time-dependent populations of selected states
$|v_1^{p_1}J_1K_1M_1\rangle^{\textrm{i}_1} |v_2^{p_2}J_2K_2M_2\rangle^{\textrm{i}_2} |n\rangle$
(panel (a)) and polaritonic states (panel (b)) as well as decomposition of polaritonic states in the $|v_1^{p_1}J_1K_1M_1\rangle^{\textrm{i}_1} |v_2^{p_2}J_2K_2M_2\rangle^{\textrm{i}_2} |n\rangle$ basis (panel (c)).
In all panels, one ortho and one para spin isomer of $^{14}$NH$_3$ are coupled to the cavity.
The notations are the same as in Fig. \ref{fig:2mol_1}.
Cavity parameters are chosen as $\omega_\textrm{c}=930.786 ~ \textrm{cm}^{-1}$, $g=10^{-4} ~ \textrm{au}$ and $\tau = 5 ~ \textrm{ps}$.
Here, mainly the ortho isomer is involved in polariton formation, however, due to weak mixings, cavity-mediated energy exchange can occur between the ortho and para isomers.}
\end{figure}

Time-dependent populations up to $t = 50 ~ \textrm{ps}$ and characterization of selected polaritonic states for two $^{14}$NH$_3$ molecules are shown in Fig. \ref{fig:2mol_1}, Fig. \ref{fig:2mol_2} and Fig. \ref{fig:2mol_3}.
In each case, cavity parameters  $g=10^{-4} ~ \textrm{au}$ and $\tau = 5 ~ \textrm{ps}$ are used.
Since we wish to focus on the impact of different nuclear spin isomers on rovibrational polaritons,
one ortho and one para spin isomer are coupled to the cavity.
We note that in the ortho-ortho and para-para cases, both molecules are expected to form polaritons.
However, in the ortho-para case, we will show that polariton formation can involve one or two molecules, depending on the cavity parameters.
The initial state of the cavity-molecule system is a pure state, that is, $\hat{\rho}_0 = |\psi_0\rangle\langle\psi_0|$ and the Lindblad equation is solved numerically according to Section \ref{sec:tdqm}.
In the following, we implicitly assume that molecular eigenstates $|v^{p}JKM \rangle$ are combined with spin functions according to the Pauli principle.
In order to improve the lucidity of our notation in this section, molecular eigenstates are labeled $|v^{p}JKM \rangle^{\textrm{i}}$ where $\textrm{i}=\textrm{p}$ and $\textrm{i}=\textrm{o}$ for the para and ortho spin isomers, respectively.
Rovibrational energy levels of $^{14}$NH$_3$ relevant for the forthcoming discussion are summarized in Table \ref{tbl:levels}.
As already stated, rovibrational energy levels computed with GENIUSH show excellent agreement with the corresponding experimental values extracted from Ref. \citenum{25SmYuTe}.

\begin{table}
  \caption{\label{tbl:levels}
  Selected rovibrational energy levels of the $^{14}$NH$_3$ molecule labeled with vibrational quantum number of the inversion mode ($v^p$, where $p=\pm$ is the vibrational parity), symmetric top quantum numbers ($J$, $K$ and $M$) and nuclear spin quantum number ($I$).
  $I=3/2$ and $I=1/2$ correspond to the ortho and para spin isomers of $^{14}$NH$_3$, respectively.
  Pauli-forbidden energy levels are explicitly indicated in the table.
  The zero-point vibrational energy equals $7430.282 ~ \textrm{cm}^{-1}$.}
  \begin{tabular}{c|c|c|c|c|c|c}
      ~~ $v^p$ ~~  & ~~ $J$ ~~  & ~~ $K$ ~~ & ~~ $M$ ~~  & ~~ $I$ ~~ & ~~ $E~/~\textrm{cm}^{-1}$ ~~ & ~~ spin isomer ~~  \\
      \hline
      $0^+$ & $1$ & $0$ & $0$ & $3/2$ & $7450.172$ & ortho   \\
      $1^-$ & $2$ & $0$ & $0$ & $3/2$ & $8457.715$ & ortho   \\
      $0^+$ & $1$ & $1$ & $0$ & $1/2$ & $7446.455$ & para \\
      $1^-$ & $2$ & $1$ & $0$ & $1/2$ & $8453.992$ & para \\
      \hline
      $0^+$ & $2$ & $0$ & $0$ & $-$ & $7489.932$ & forbidden \\
      $1^-$ & $1$ & $0$ & $0$ & $-$ & $8418.184$ & forbidden \\
      $0^-$ & $2$ & $0$ & $0$ & $3/2$ & $7490.694$ & ortho \\
      $1^+$ & $1$ & $0$ & $0$ & $3/2$ & $8382.895$ & ortho \\
      $0^+$ & $2$ & $1$ & $0$ & $1/2$ & $7486.221$ & para \\
      $1^-$ & $1$ & $1$ & $0$ & $1/2$ & $8414.457$ & para \\
      \hline
      $0^-$ & $3$ & $3$ & $1$ & $3/2$ & $7516.941$ & ortho   \\
      $1^+$ & $3$ & $3$ & $1$ & $3/2$ & $8447.727$ & ortho   \\
      $0^-$ & $3$ & $2$ & $1$ & $1/2$ & $7535.466$ & para \\
      $1^+$ & $3$ & $2$ & $1$ & $1/2$ & $8467.581$ & para \\
  \end{tabular}
\end{table}

In Fig. \ref{fig:2mol_1}, the cavity wavenumber is chosen as $\omega_\textrm{c} = 1007.543 ~ \textrm{cm}^{-1}$ and the initial state equals
 $|\psi_0\rangle = |v_1^{p_1}J_1K_1M_1 \rangle^{\textrm{i}_1} |v_2^{p_2}J_2K_2M_2 \rangle^{\textrm{i}_2} |n\rangle = 
 |0^+,1,0,0 \rangle^{\textrm{o}} |0^+,1,1,0 \rangle^{\textrm{p}} |1\rangle$. 
In this case, $\omega_\textrm{c}$ is resonant with the ortho transition
$|0^+,1,0,0 \rangle^{\textrm{o}} \leftrightarrow |1^-,2,0,0 \rangle^{\textrm{o}}$ 
($K=0$, transition wavenumber is $\omega_\textrm{o} = 1007.543 ~ \textrm{cm}^{-1}$)
and $\omega_\textrm{c}$ is in near resonance with the para transition
$|0^+,1,1,0 \rangle^{\textrm{p}} \leftrightarrow |1^-,2,1,0 \rangle^{\textrm{p}}$ 
($K=1$, transition wavenumber is $\omega_\textrm{p} = 1007.537 ~ \textrm{cm}^{-1}$).
Both of these transitions are dipole-allowed, which leads to the formation of collective ortho-para rovibrational polaritons for the cavity parameters used.
We note that the difference 
$\omega_\textrm{o}-\omega_\textrm{p} = 0.006~ \textrm{cm}^{-1}$
agrees well with its experimental counterpart
of $0.006602 ~ \textrm{cm}^{-1}$.\cite{25SmYuTe}
Since $|\psi_0\rangle$ is not an eigenstate of the cavity-molecule system, one can observe Rabi oscillations between the states
$|0^+,1,0,0 \rangle^{\textrm{o}} |0^+,1,1,0 \rangle^{\textrm{p}} |1\rangle$, 
$|1^-,2,0,0 \rangle^{\textrm{o}} |0^+,1,1,0 \rangle^{\textrm{p}} |0\rangle$ and
$|0^+,1,0,0 \rangle^{\textrm{o}} |1^-,2,1,0 \rangle^{\textrm{p}} |0\rangle$
in panel (a) of Fig. \ref{fig:2mol_1}, implying energy exchange between the molecules and the cavity mode.
In addition, due to the finite lifetime of cavity excitations, the state 
$|0^+,1,0,0 \rangle^{\textrm{o}} |0^+,1,1,0 \rangle^{\textrm{p}} |1\rangle$
decays to 
$|0^+,1,0,0 \rangle^{\textrm{o}} |0^+,1,1,0 \rangle^{\textrm{p}} |0\rangle$.
It is also visible in panel (a) that the ortho and para population transfers occur with slightly different amplitudes due to differences in dipole matrix elements.

Panels (b) and (c) of Fig. \ref{fig:2mol_1} show time-dependent populations as well as the characterization of polaritonic states (eigenstates of the coupled cavity-molecule system) relevant for the dynamics.
Each polaritonic state in Fig. \ref{fig:2mol_1} is decomposed as
\begin{equation}
    |\Psi_k\rangle = \sum_{v_1p_1J_1} \sum_{v_2p_2J_2} \sum_n c^{(k)}_{v_1 p_1 J_1 v_2 p_2 J_2 n}
    |v_1^{p_1}J_1K_1M_1 \rangle^{\textrm{i}_1} |v_2^{p_2}J_2K_2M_2 \rangle^{\textrm{i}_2} |n\rangle
\end{equation}
and the absolute squares
$|c^{(k)}_{v_1 p_1 J_1 v_2 p_2 J_2 n}|^2$
are depicted in panel (c) of Fig. \ref{fig:2mol_1} for the most significant contributions.
In Fig. \ref{fig:2mol_1}, energies $E_k$ of selected polaritonic states are referenced to the energy of the lowest polaritonic state.
It is visible in panel (b) that polaritonic states $|\Psi_1\rangle$ and $|\Psi_3\rangle$
($E_1=1025.171 ~ \textrm{cm}^{-1}$ and $E_3=1028.047 ~ \textrm{cm}^{-1}$)
carry appreciable populations in the singly-excited manifold, while
polaritonic state $|\Psi_2\rangle$ with 
$E_2=1026.581 ~ \textrm{cm}^{-1}$
is not involved in the dynamics.
In addition, the polaritonic state with $E = 19.056 ~ \textrm{cm}^{-1}$ is populated by the incoherent decay term of the Lindblad equation and corresponds to the state
$|0^+,1,0,0 \rangle^{\textrm{o}} |0^+,1,1,0 \rangle^{\textrm{p}} |0\rangle$
to a very good approximation.
Panel (c) shows that $|\Psi_2\rangle$ gives zero overlap with Fock state $|1\rangle$, in contrast to polaritonic states $|\Psi_1\rangle$ and $|\Psi_3\rangle$.
The qualitative interpretation of these results is facilitated by the Tavis--Cummings model using the cavity parameters of Fig. \ref{fig:2mol_1}.
Assuming resonant cavity-molecule coupling, the bright Hamiltonian defined in Eq. \eqref{eq:mx_bright} can be approximated as 
\begin{equation}
    \mathbf{H}_\textrm{bright}^{(1)} =
    \left[
    \begin{array}{ccc}
            0 & V & V \\
            V & 0 & 0 \\
            V & 0 & 0
        \end{array}
    \right]
    \label{eq:mx_bright_2mol_1}
\end{equation}
since $n_\textrm{o} = n_\textrm{p} = 1$,
$\omega_\textrm{o} \approx \omega_\textrm{p}$,
the detunings are zero (or negligible) and
$V_\textrm{o} \approx V_\textrm{p} \approx V$ in this particular case.
Under these assumptions, the three orthonormal eigenvectors of $\mathbf{H}_\textrm{bright}^{(1)}$ correspond to a dark state
(see also the text below Eq. \eqref{eq:bright_basis})
\begin{equation}
    \mathbf{v}_0^{(1)} = 
        \frac{1}{\sqrt{2}}
        \left(
            \begin{array}{c}
                0 \\
                1 \\
                -1
            \end{array}
        \right)
    \label{eq:fig1dark}
\end{equation}
with an eigenvalue of $\lambda_0^{(1)} = 0$ and two bright states
\begin{equation}
    \mathbf{v}_\pm^{(1)} = 
        \frac{1}{2}
        \left(
            \begin{array}{c}
                \pm \sqrt{2} \\
                1 \\
                1
            \end{array}
        \right)
    \label{eq:fig1bright}
\end{equation}
with $\lambda_\pm^{(1)} = \pm \sqrt{2} V$.
The order of basis states corresponds to
$|0^+,1,0,0 \rangle^{\textrm{o}} |0^+,1,1,0 \rangle^{\textrm{p}} |1\rangle$, 
$|1^-,2,0,0 \rangle^{\textrm{o}} |0^+,1,1,0 \rangle^{\textrm{p}} |0\rangle$ and
$|0^+,1,0,0 \rangle^{\textrm{o}} |1^-,2,1,0 \rangle^{\textrm{p}} |0\rangle$
in the previous equations.
Clearly, eigenvectors $\mathbf{v}_0^{(1)}$ and $\mathbf{v}_\pm^{(1)}$ represent collective ortho-para polaritons.
Comparing Eq. \eqref{eq:fig1dark} and Eq. \eqref{eq:fig1bright} to panel (c) of Fig. \ref{fig:2mol_1} reveals that the dark eigenvector $\mathbf{v}_0^{(1)}$ can be assigned to $|\Psi_2\rangle$, while bright eigenvectors $\mathbf{v}_\pm^{(1)}$ correspond to $|\Psi_1\rangle$ and $|\Psi_3\rangle$, respectively.
The structure of $\mathbf{H}_\textrm{bright}^{(1)}$ implies that $|\Psi_2\rangle$ becomes essentially dark and is not involved in the dynamics due to virtually zero overlap with the initial state 
$|\psi_0\rangle = |0^+,1,0,0 \rangle^{\textrm{o}} |0^+,1,1,0 \rangle^{\textrm{p}} |1\rangle$.

In Fig. \ref{fig:2mol_2}, the cavity wavenumber is set to $\omega_\textrm{c} = 928.236 ~ \textrm{cm}^{-1}$ which is resonant with the para transition $|0^+,2,1,0 \rangle^{\textrm{p}} \leftrightarrow |1^-,1,1,0 \rangle^{\textrm{p}}$ having a transition wavenumber of $\omega_\textrm{p} = 928.236 ~ \textrm{cm}^{-1}$.
Disregarding the Pauli principle, one could argue that there is a close-lying ortho transition at $\omega_\textrm{o} = 928.253 ~ \textrm{cm}^{-1}$ linking rovibrational eigenstates $|0^+,2,0,0 \rangle^{\textrm{o}}$ and $|1^-,1,0,0 \rangle^{\textrm{o}}$.
However, according to Table \ref{tbl:pauli}, the eigenstates  $|0^+,2,0,0 \rangle^{\textrm{o}}$ and $|1^-,1,0,0 \rangle^{\textrm{o}}$ are both Pauli-forbidden for $K=0$.
Therefore, we assume that the ortho molecule is initially prepared in the Pauli-allowed state $|0^-,2,0,0 \rangle^{\textrm{o}}$ instead of $|0^+,2,0,0 \rangle^{\textrm{o}}$.
This choice is motivated by the fact that $|0^-,2,0,0 \rangle^{\textrm{o}}$ is the closest to $|0^+,2,0,0 \rangle^{\textrm{o}}$ in energy with a separation of $0.762 ~ \textrm{cm}^{-1}$.
The state $|0^-,2,0,0 \rangle^{\textrm{o}}$ has a dipole-allowed transition to $|1^+,1,0,0 \rangle^{\textrm{o}}$ at $\omega_\textrm{o} = 892.201 ~ \textrm{cm}^{-1}$, far off-resonant with $\omega_\textrm{c} = 928.236 ~ \textrm{cm}^{-1}$.
Therefore, if the initial state equals
$|\psi_0\rangle = |0^-,2,0,0 \rangle^{\textrm{o}} |0^+,2,1,0 \rangle^{\textrm{p}} |1\rangle$,
only the Rabi oscillation between states
$|0^-,2,0,0 \rangle^{\textrm{o}} |0^+,2,1,0 \rangle^{\textrm{p}} |1\rangle$ and $|0^-,2,0,0 \rangle^{\textrm{o}} |1^-,1,1,0 \rangle^{\textrm{p}} |0\rangle$ takes place as shown in panel (a) of Fig. \ref{fig:2mol_2}.
Due to the Pauli principle, the ortho isomer acts as a spectator and is not involved in polariton formation.
Therefore, results in Fig. \ref{fig:2mol_2} are essentially identical to the single-molecule para case with initial state $|\psi_0\rangle = |0^+,2,1,0 \rangle^{\textrm{p}} |1\rangle$.
It is striking in Fig. \ref{fig:2mol_2} that the oscillation period in panel (a) is increased compared to panel (a) of Fig. \ref{fig:2mol_1}.
In other words, collective coupling is reduced to the single-molecule level in Fig. \ref{fig:2mol_2}.
This is a clear indication that collective coupling effects are indeed modified by the Pauli principle and nuclear spin degrees of freedom.

Polaritonic states are characterized by panels (b) and (c) of Fig. \ref{fig:2mol_2}.
According to panel (b), polaritonic states $|\Psi_1\rangle$ and $|\Psi_2\rangle$ are populated initially in the singly-excited manifold with $E_1=1026.634 ~ \textrm{cm}^{-1}$ and $E_2=1028.518 ~ \textrm{cm}^{-1}$.
Moreover, polaritonic state with $E=99.313 ~ \textrm{cm}^{-1}$, again populated by the incoherent decay term, can essentially be assigned to the state
$|0^-,2,0,0 \rangle^{\textrm{o}} |0^+,2,1,0 \rangle^{\textrm{p}} |0\rangle$.
Panel (c) shows that both $|\Psi_1\rangle$ and $|\Psi_2\rangle$ are superpositions of basis states
$|0^-,2,0,0 \rangle^{\textrm{o}} |0^+,2,1,0 \rangle^{\textrm{p}} |1\rangle$ and 
$|0^-,2,0,0 \rangle^{\textrm{o}} |1^-,1,1,0 \rangle^{\textrm{p}} |0\rangle$.
These results can be understood qualitatively using the bright Hamiltonian of Eq. \eqref{eq:mx_bright} which has now the form
\begin{equation}
    \mathbf{H}_\textrm{bright}^{(2)} =
    \left[
    \begin{array}{ccc}
            0 & V_\textrm{o} & V_\textrm{p} \\
            V_\textrm{o} & \Delta_\textrm{o} & 0 \\
            V_\textrm{p} & 0 & 0
        \end{array}
    \right]
    \label{eq:mx_bright_2mol_2}
\end{equation}
with $\Delta_\textrm{p} = 0$.
Since $|\Delta_\textrm{o}| >> |V_\textrm{o}|$, the ortho isomer is effectively decoupled and the three orthonormal eigenvectors of $\mathbf{H}_\textrm{bright}^{(2)}$ can be very well approximated by an ortho dark state
\begin{equation}
    \mathbf{v}_0^{(2)} = 
        \left(
            \begin{array}{c}
                0 \\
                1 \\
                0
            \end{array}
        \right)
    \label{eq:fig2dark}
\end{equation}
with eigenvalue $\lambda_0^{(2)} = \Delta_\textrm{o}$ and two para bright states
\begin{equation}
    \mathbf{v}_\pm^{(2)} = 
        \frac{1}{\sqrt{2}}
        \left(
            \begin{array}{c}
                1 \\
                0 \\
                \pm 1
            \end{array}
        \right)
    \label{eq:fig2bright}
\end{equation}
with $\lambda_\pm^{(2)} = \pm V_\textrm{p}$.
Here, the order of the three basis states is 
$|0^-,2,0,0 \rangle^{\textrm{o}} |0^+,2,1,0 \rangle^{\textrm{p}} |1\rangle$,
$|1^+,1,0,0 \rangle^{\textrm{o}} |0^+,2,1,0 \rangle^{\textrm{p}} |0\rangle$
and 
$|0^-,2,0,0 \rangle^{\textrm{o}} |1^-,1,1,0 \rangle^{\textrm{p}} |0\rangle$.
According to panel (c) of Fig. \ref{fig:2mol_2}, polaritonic states $|\Psi_1\rangle$ and $|\Psi_2\rangle$ can be assigned to para bright states $\mathbf{v}_\pm^{(2)}$.

So far, the cavity mode has initially occupied Fock state $|1\rangle$.
In Fig. \ref{fig:2mol_3}, results are shown for the initial state
$|\psi_0\rangle = |1^+,3,3,1 \rangle^{\textrm{o}} |0^-,3,2,1 \rangle^{\textrm{p}} |0\rangle$.
Now, $\omega_\textrm{c} = 930.786 ~ \textrm{cm}^{-1}$ is resonant with the ortho transition $|1^+,3,3,1 \rangle^{\textrm{o}} \leftrightarrow |0^-,3,3,1 \rangle^{\textrm{o}}$ ($K=3$, transition wavenumber is $\omega_\textrm{o} = 930.786 ~ \textrm{cm}^{-1}$).
Thus, as shown in panel (a) of Fig. \ref{fig:2mol_3}, energy is transferred from the ortho isomer to the cavity mode due to resonant cavity-molecule coupling.
Note that the para isomer can also absorb some energy as the transition wavenumber $\omega_\textrm{p}=932.115 ~ \textrm{cm}^{-1}$ of the para transition
$|0^-,3,2,1 \rangle^{\textrm{p}} \leftrightarrow |1^+,3,2,1 \rangle^{\textrm{p}}$
($K=2$) is still not completely out of resonance with $\omega_\textrm{c}$ for $g=10^{-4} ~ \textrm{au}$.
In this case, mainly the ortho isomer is involved in polariton formation, however, weak admixtures from the para isomer lead to cavity-mediated ortho-para intermolecular energy transfer.
Similarly to the previous cases, panels (b) and (c) of Fig. \ref{fig:2mol_3} show time-dependent polaritonic populations and characterize polaritonic states relevant for the dynamics.
Now, mainly polaritonic states $|\Psi_1\rangle$ and $|\Psi_2\rangle$ ($E_1=991.367 ~ \textrm{cm}^{-1}$ and $E_2=992.427 ~ \textrm{cm}^{-1}$) are populated in the singly-excited manifold, while polaritonic state $|\Psi_3\rangle$ with $E_3=993.402 ~ \textrm{cm}^{-1}$ is also populated to some extent.
This observation is in line with panel (c) which shows that $|\Psi_1\rangle$ and $|\Psi_2\rangle$ are largely bright ortho polaritons with small contribution from the state
$|0^-,3,3,1 \rangle^{\textrm{o}} |1^+,3,2,1 \rangle^{\textrm{p}} |0\rangle$.
At the same time, $|\Psi_3\rangle$ mainly involves basis state 
$|0^-,3,3,1 \rangle^{\textrm{o}} |1^+,3,2,1 \rangle^{\textrm{p}} |0\rangle$
with a small photonic component.
Finally, the polaritonic state with $E=61.164 ~ \textrm{cm}^{-1}$ is essentially identical to state $|0^-,3,3,1 \rangle^{\textrm{o}} |0^-,3,2,1 \rangle^{\textrm{p}} |0\rangle$.
The numerical results can be further rationalized by considering the bright Hamiltonian which becomes
\begin{equation}
    \mathbf{H}_\textrm{bright}^{(3)} =
    \left[
    \begin{array}{ccc}
            0 & V_\textrm{o} & V_\textrm{p} \\
            V_\textrm{o} & 0 & 0 \\
            V_\textrm{p} & 0 & \Delta_\textrm{p}
        \end{array}
    \right]
    \label{eq:mx_bright_2mol_3}
\end{equation}
with $\Delta_\textrm{o} = 0$.
In contrast to the cases presented in Fig. \ref{fig:2mol_1} and Fig. \ref{fig:2mol_2}, $\mathbf{H}_\textrm{bright}^{(3)}$ cannot be diagonalized by simple analytical considerations.
Finally, we note that the DSE and polarizability terms have a negligible effect for the cavity parameters used in this section.

\section{Ensemble results}

Next, we examine an ortho-para ensemble of $^{14}$NH$_3$ molecules at a given temperature $T$ and simulate cavity transmission spectra using the Tavis--Cummings model.
Here, theoretical considerations presented in Section \ref{sec:tc} and Section \ref{sec:transspec} are utilized, and the DSE and polarizability terms are neglected.
In order to relate the current results to recent gas-phase rovibrational polaritonic experiments for the CH$_4$ molecule,\cite{23WrNeWe,23WrNeWe_2,24NeWe}
we adopt the experimental parameters of Ref. \citenum{23WrNeWe} where a temperature of $T = 120 ~ \textrm{K}$ and a mode volume of $V_\textrm{exp} = 7.7 \cdot 10^{-3} ~ \textrm{cm}^3$ were reported.
In addition, the number of CH$_4$ molecules in the cavity was estimated to be $n_\textrm{exp} = 2.7 \cdot 10^{13}$ at the highest density.
Using the relation $g = \sqrt{\frac{\hbar \omega_\textrm{c}}{2 \epsilon_0 V_\textrm{exp}}}$ and the cavity frequencies
$\omega_\textrm{c} \approx 1000 ~ \textrm{cm}^{-1}$
applied in Section \ref{sec:2molres}, the coupling strength parameter is on the order of $g = 10^{-12} ~ \textrm{au}$, which will be used throughout this section.
The transmission spectrum is modeled as the sum of Lorentzian peaks
\begin{equation}
    f(\omega) = \frac{I}{\pi} \frac{\varGamma/2}{(\omega-\omega_0)^2 + (\varGamma/2)^2}
\label{eq:lorentz}
\end{equation}
centered at $\omega_0$ with intensity $I$ and full width at half-maximum $\varGamma$.
Here, we choose 
$\varGamma = 65 ~ \textrm{MHz} \approx 2.17 \cdot 10^{-3} ~ \textrm{cm}^{-1}$
which corresponds to the cavity linewidth of Ref. \citenum{23WrNeWe}.
In contrast to the method introduced in Ref. \citenum{24YuKo}, our approach does not directly provide the transmission lineshape function.
Therefore, as an approximation, we employ Eq. \eqref{eq:lorentz} to model the transmission spectrum and use the same width for each spectrum line.

\begin{figure}
\includegraphics[scale=0.5]{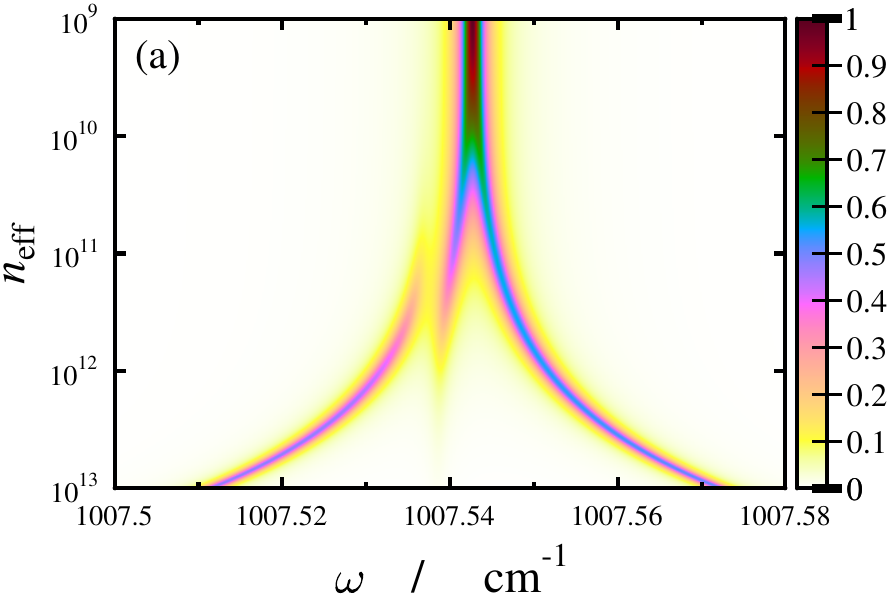}
\includegraphics[scale=0.5]{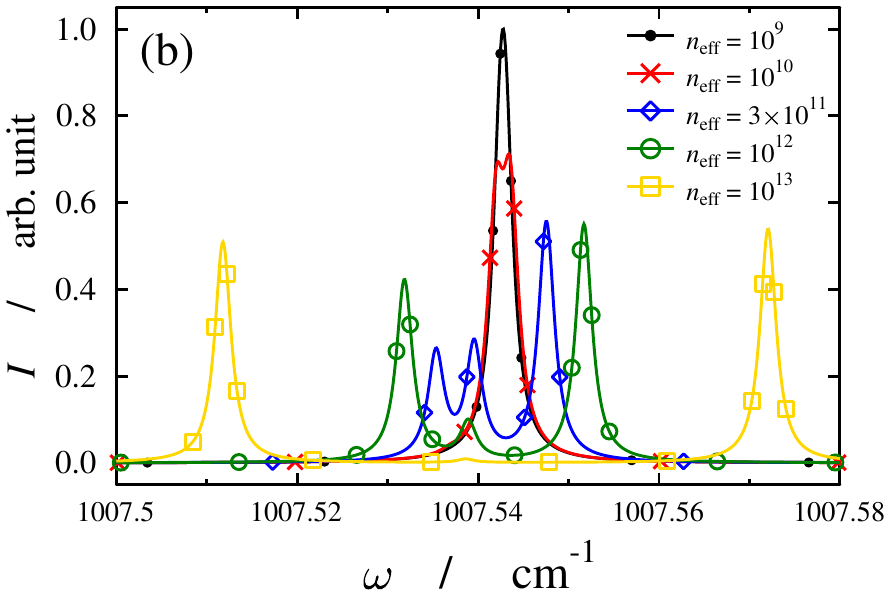}
\caption{\label{fig:nmol_1}
Cavity transmission spectrum of an ortho-para ensemble of $^{14}$NH$_3$ molecules ($\omega$ denotes the transition wavenumber in units of $\textrm{cm}^{-1}$, while $I$ is the intensity in arbitrary units).
Panel (a) shows the spectrum as a function of the number of molecules $n_\textrm{eff}$, while panel (b) gives the spectrum for selected values of $n_\textrm{eff}$.
Cavity parameters are chosen as 
$\omega_\textrm{c} = 1007.543 ~ \textrm{cm}^{-1}$
and $g = 10^{-12} ~ \textrm{au}$, and the spectrum is modeled as the sum of Lorentzian peaks with full width at half-maximum
$\varGamma = 65 ~ \textrm{MHz}$.
In this case, the cavity can couple to close-lying ortho
($\omega_\textrm{o} = 1007.543 ~ \textrm{cm}^{-1}$)
and para 
($\omega_\textrm{p} = 1007.537 ~ \textrm{cm}^{-1}$)
transitions and the cavity frequency is resonant with the ortho transition.
For smaller values of $n_\textrm{eff}$, only the ortho molecules are involved in polariton formation, while for larger values of $n_\textrm{eff}$, ortho-para polaritons are formed.}
\end{figure}

In Fig. \ref{fig:nmol_1}, the close-lying ortho and para transitions 
$|0^+,1,0,0 \rangle^{\textrm{o}} \leftrightarrow |1^-,2,0,0 \rangle^{\textrm{o}}$
and
$|0^+,1,1,0 \rangle^{\textrm{p}} \leftrightarrow |1^-,2,1,0 \rangle^{\textrm{p}}$ 
are considered, similar to the two-molecule results presented in Fig. \ref{fig:2mol_1}.
In this case, the respective energy levels are
$E_\textrm{o}^{(g)} = 7450.172 ~ \textrm{cm}^{-1}$,
$E_\textrm{o}^{(e)} = 8457.715 ~ \textrm{cm}^{-1}$,
$E_\textrm{p}^{(g)} = 7446.455 ~ \textrm{cm}^{-1}$
and
$E_\textrm{p}^{(e)} = 8453.992 ~ \textrm{cm}^{-1}$,
while the transition wavenumbers equal
$\omega_\textrm{o} = 1007.543 ~ \textrm{cm}^{-1}$ (ortho)
and
$\omega_\textrm{p} = 1007.537 ~ \textrm{cm}^{-1}$ (para).
Using the experimental energy levels and transition frequencies of Ref. \citenum{25SmYuTe}, we have verified that the close-lying ortho and para transitions selected here can be safely isolated from other transitions.
Fig. \ref{fig:nmol_1} shows cavity transmission spectra for resonant cavity coupling with the ortho transition, that is,
$\omega_\textrm{c} = \omega_\textrm{o}$ ($\Delta_\textrm{o}=0$).
In panel (a) of Fig. \ref{fig:nmol_1}, the coupling strength is kept fixed at $g = 10^{-12} ~ \textrm{au}$ and the effective number of molecules is varied between
$n_\textrm{eff} = 10^9$ and $n_\textrm{eff} = 10^{13}$.
Here, $n_\textrm{eff}$ gives the total number of molecules occupying the eigenstates
$|0^+,1,0,0 \rangle^{\textrm{o}}$ and $|0^+,1,1,0 \rangle^{\textrm{p}}$.
The values of $n_\textrm{eff}$ used in this work were devised based on the maximum number of molecules $n_\textrm{exp}=2.7 \cdot 10^{13}$ reported in Ref. \citenum{23WrNeWe} and by estimating the relevant Boltzmann factors for $T = 120 ~ \textrm{K}$.
Panel (b) of Fig. \ref{fig:nmol_1} provides cavity transmission spectra for selected values of $n_\textrm{eff}$.

For the smallest value of $n_\textrm{eff} = 10^9$ in Fig. \ref{fig:nmol_1},
the spectrum features a single peak that splits into two peaks as $n_\textrm{eff}$ increases, indicating the onset of the strong coupling regime.
Initially, only the ortho molecules couple to the cavity field and ortho polaritons are formed, while the para molecules act as spectators.
This case corresponds to a simple Rabi splitting and the separation of the peaks ($2v_{\textrm{o},\kappa}$) is solely determined by the ortho collective coupling matrix element $v_{\textrm{o},\kappa}$ defined in Eq. \eqref{eq:collective_coupling}.
For higher values of $n_\textrm{eff}$, three peaks appear in the spectrum, which can be attributed to the emergence of three bright ortho-para polaritons, each carrying nonzero photonic component.
Finally, at the highest values of $n_\textrm{eff}$ used in this work, one of the three bright polaritons becomes essentially dark, which can be explained by considerations outlined below Eq. \eqref{eq:bright_basis} and Eq. \eqref{eq:mx_bright_2mol_1} for the two-molecule case.
In this case, the spectrum again features two peaks whose separation is now given by
$2\sqrt{v_{\textrm{o},\kappa}^2+v_{\textrm{p},\lambda}^2}$ to a good approximation.

\begin{figure}
\includegraphics[scale=0.5]{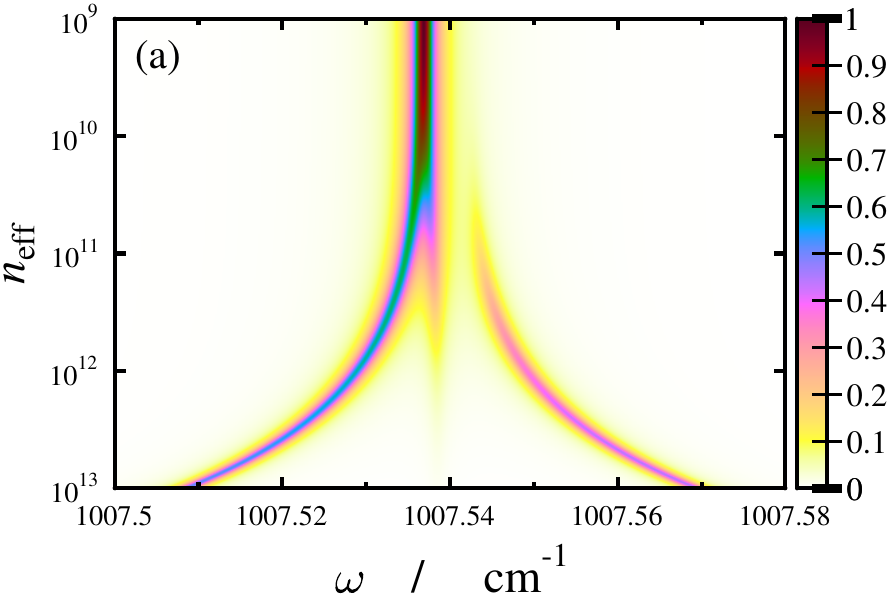}
\includegraphics[scale=0.5]{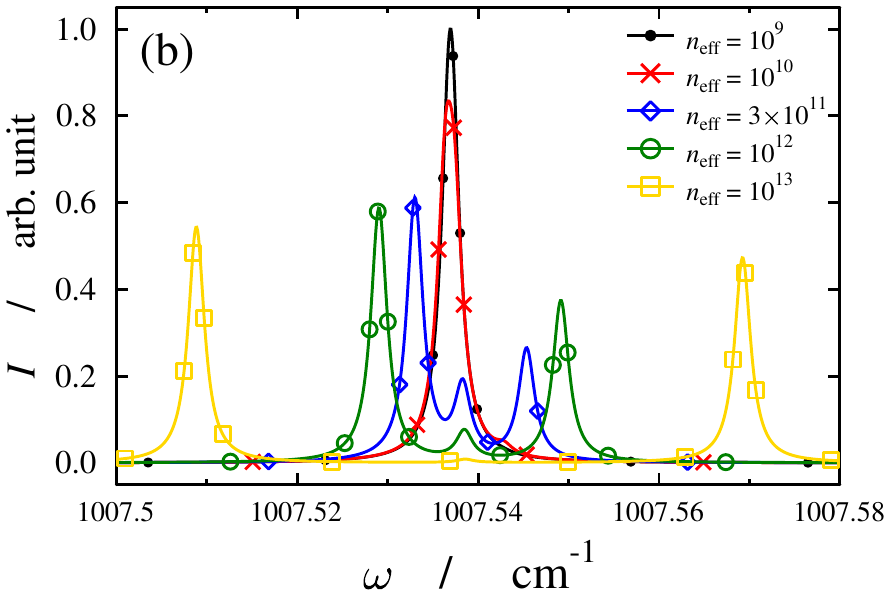}
\caption{\label{fig:nmol_2}
Cavity transmission spectrum of an ortho-para ensemble of $^{14}$NH$_3$ molecules.
Panel (a) shows the spectrum as a function of the number of molecules $n_\textrm{eff}$, while panel (b) gives the spectrum for selected values of $n_\textrm{eff}$.
The notations are the same as in Fig. \ref{fig:nmol_1}.
Cavity parameters are chosen as 
$\omega_\textrm{c} = 1007.537 ~ \textrm{cm}^{-1}$
and $g = 10^{-12} ~ \textrm{au}$, and the spectrum is modeled as the sum of Lorentzian peaks with full width at half-maximum
$\varGamma = 65 ~ \textrm{MHz}$.
In this case, the cavity can couple to close-lying ortho
($\omega_\textrm{o} = 1007.543 ~ \textrm{cm}^{-1}$)
and para 
($\omega_\textrm{p} = 1007.537 ~ \textrm{cm}^{-1}$)
transitions and the cavity frequency is resonant with the para transition.
For smaller values of $n_\textrm{eff}$, only the para molecules are involved in polariton formation, while for larger values of $n_\textrm{eff}$, ortho-para polaritons are formed.}
\end{figure}

Fig. \ref{fig:nmol_2} pertains to the case presented in Fig. \ref{fig:nmol_1}, with the only difference being the cavity wavenumber
$\omega_\textrm{c} = 1007.537 ~ \textrm{cm}^{-1}$
which is now resonant with the para transition, that is,
$\omega_\textrm{c} = \omega_\textrm{p}$ ($\Delta_\textrm{p}=0$).
In panels (a) and (b) of Fig. \ref{fig:nmol_2} one can observe effects analogous to the ones discussed in relation to Fig. \ref{fig:nmol_1}.
Namely, for smaller values of $n_\textrm{eff}$, para polaritons are formed and two peaks can be observed in the spectrum.
As $n_\textrm{eff}$ increases, three bright ortho-para polaritons emerge, indicated by three peaks.
Finally, for the highest $n_\textrm{eff}$ values, one of the three bright polaritons becomes virtually dark and the spectrum again features two peaks.

\begin{figure}
\includegraphics[scale=0.5]{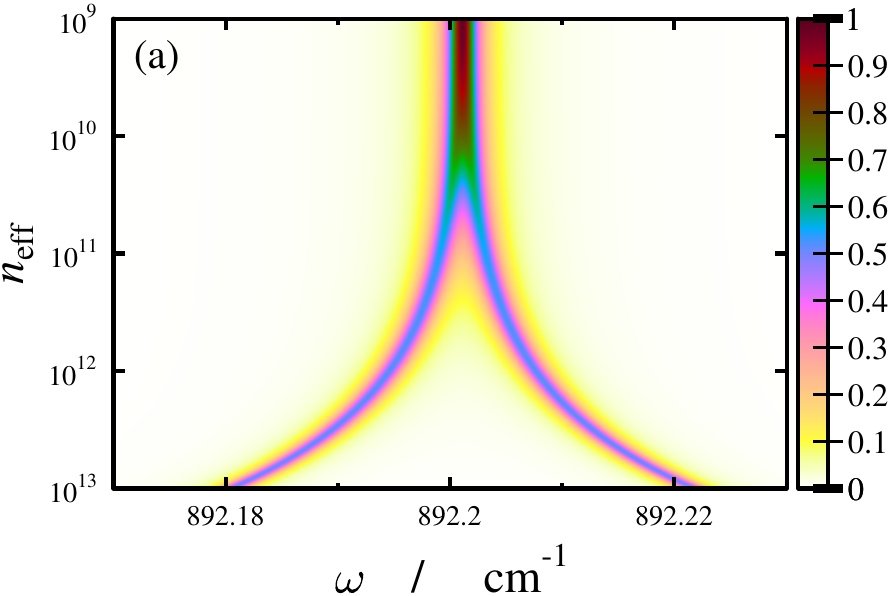}
\includegraphics[scale=0.5]{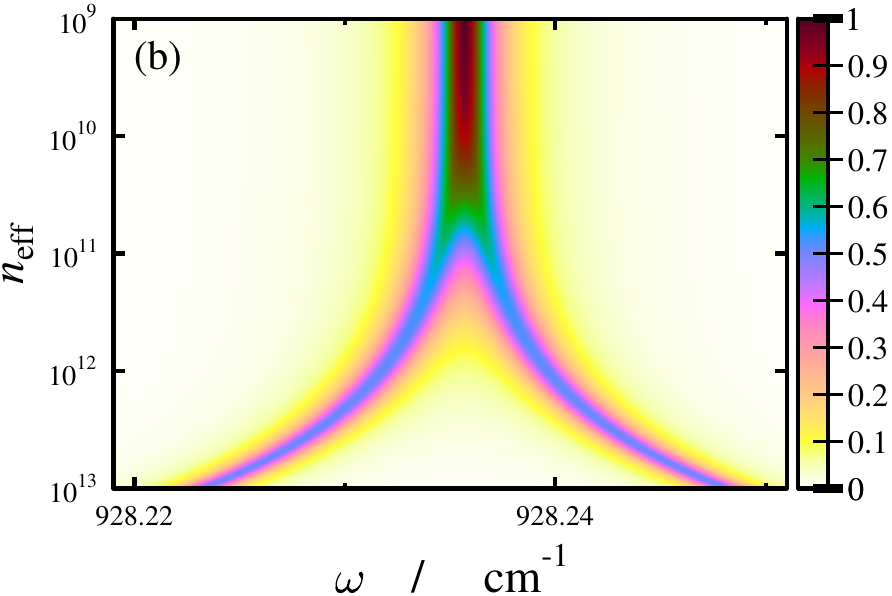}
\caption{\label{fig:nmol_3}
Cavity transmission spectrum of an ortho-para ensemble of $^{14}$NH$_3$ molecules as a function of the number of molecules $n_\textrm{eff}$ ($\omega$ denotes the transition wavenumber in units of $\textrm{cm}^{-1}$, while $I$ is the intensity in arbitrary units).
The cavity wavenumber is chosen as 
$\omega_\textrm{c} = 892.201 ~ \textrm{cm}^{-1}$ (panel (a), resonant with the ortho transition) and
$\omega_\textrm{c} = 928.236 ~ \textrm{cm}^{-1}$ (panel (b), resonant with the para transition).
In addition, $g = 10^{-12} ~ \textrm{au}$ and the spectrum is modeled as the sum of Lorentzian peaks with full width at half-maximum
$\varGamma = 65 ~ \textrm{MHz}$.
In this case, the ortho and para transitions are well isolated, and the cavity couples selectively to the ortho (panel (a)) or para (panel (b)) molecules, regardless of $n_\textrm{eff}$.}
\end{figure}

In contrast to the close-lying ortho and para transitions of Fig. \ref{fig:nmol_1} and Fig. \ref{fig:nmol_2}, Fig. \ref{fig:nmol_3} presents a scenario with well-isolated ortho and para transitions (based on two-molecule results shown in Fig. \ref{fig:2mol_2}).
In this case, the ortho and para transitions considered are
$|0^-,2,0,0 \rangle^{\textrm{o}} \leftrightarrow |1^+,1,0,0 \rangle^{\textrm{o}}$
and 
$|0^+,2,1,0 \rangle^{\textrm{p}} \leftrightarrow |1^-,1,1,0 \rangle^{\textrm{p}}$.
Therefore, the Tavis--Cummings model is now parameterized with energy levels
$E_\textrm{o}^{(g)} = 7490.694 ~ \textrm{cm}^{-1}$,
$E_\textrm{o}^{(e)} = 8382.895 ~ \textrm{cm}^{-1}$,
$E_\textrm{p}^{(g)} = 7486.221 ~ \textrm{cm}^{-1}$
and
$E_\textrm{p}^{(e)} = 8414.457 ~ \textrm{cm}^{-1}$,
which yield the transition wavenumbers
$\omega_\textrm{o} = 892.201 ~ \textrm{cm}^{-1}$ (ortho)
and
$\omega_\textrm{p} = 928.236 ~ \textrm{cm}^{-1}$ (para).
Panels (a) and (b) show the cavity transmission spectrum as a function of $n_\textrm{eff}$ for resonant cavity coupling with the ortho and para transitions, respectively.
Again, $g=10^{-12} ~ \textrm{au}$ and $n_\textrm{eff} = 10^9-10^{13}$.
It is striking in Fig. \ref{fig:nmol_3} that for
$\omega_\textrm{c} = \omega_\textrm{o}$ ($\Delta_\textrm{o}=0$, panel (a)), 
only ortho molecules are involved in polariton formation and the separation of the two peaks in the spectrum is equal to $2v_{\textrm{o},\kappa}$.
Similarly, if $\omega_\textrm{c} = \omega_\textrm{p}$ ($\Delta_\textrm{p}=0$, panel (b)), two para polaritons emerge and the Rabi splitting is $2v_{\textrm{p},\lambda}$.
Depending on $\omega_\textrm{c}$, even for the highest value of $n_\textrm{eff}$, either the para or the ortho molecules are spectators
due to the lack of dipole and Pauli-allowed transitions that are resonant with the cavity frequency.
Consequently, the collective coupling strength is reduced compared to what one would naively expect for $n_\textrm{eff}$ molecules. 

\begin{figure}
\includegraphics[scale=0.5]{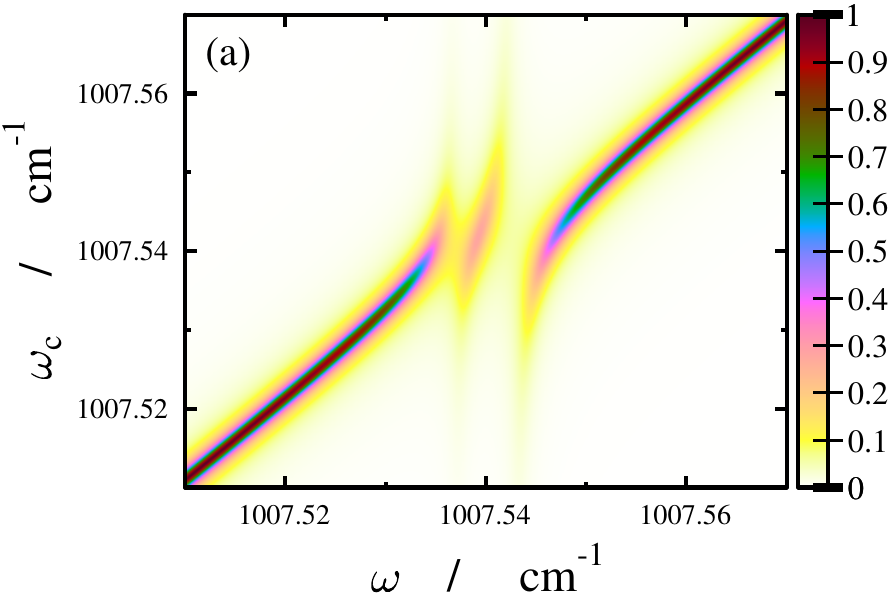}
\includegraphics[scale=0.5]{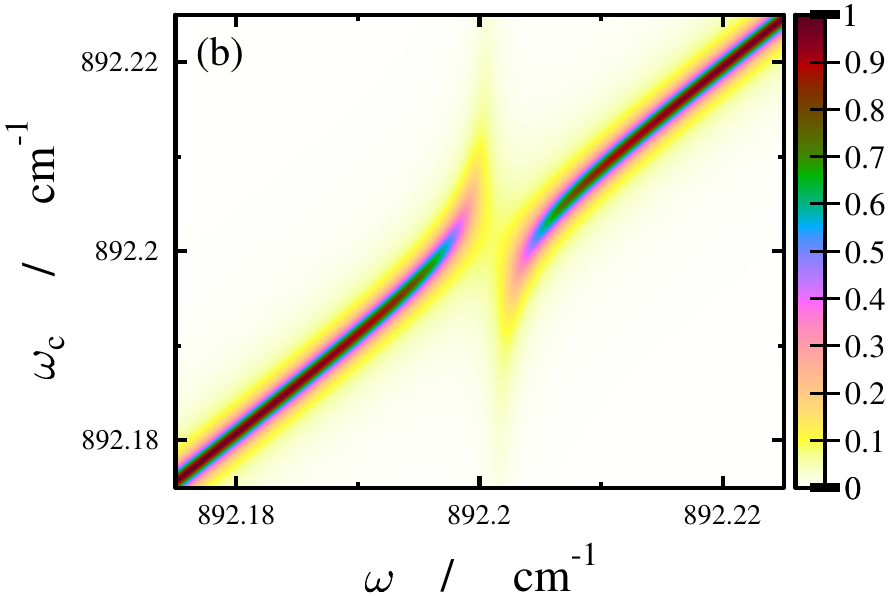}
\caption{\label{fig:dispersion}
Cavity transmission spectrum of an ortho-para ensemble of $^{14}$NH$_3$ molecules as a function of the cavity wavenumber $\omega_\textrm{c}$ ($\omega$ denotes the transition wavenumber in units of $\textrm{cm}^{-1}$, while $I$ is the intensity in arbitrary units).
In panel (a), the cavity can simultaneously couple to the ortho and para transitions (using the scenario shown in Fig. \ref{fig:nmol_1} and Fig. \ref{fig:nmol_2}),
while in panel (b), only the ortho molecules are involved in the formation of polaritons (see panel (a) of Fig. \ref{fig:nmol_3}).
The number of molecules equals 
$n_\textrm{eff} = 3 \cdot 10^{11}$, 
$g = 10^{-12} ~ \textrm{au}$ and the spectrum is modeled as the sum of Lorentzian peaks with full width at half-maximum
$\varGamma = 65 ~ \textrm{MHz}$.}
\end{figure}

Up to this point, we have always assumed resonant cavity-molecule coupling.
Now, the cavity frequency is systematically detuned from resonance and the resulting dispersion plots are shown in Fig. \ref{fig:dispersion} with $g = 10^{-12} ~ \textrm{au}$ and
$n_\textrm{eff} = 3 \cdot 10^{11}$.
In panels (a) and (b) of Fig. \ref{fig:dispersion}, the cavity frequency is swept across the close-lying ortho and para resonances of Fig. \ref{fig:nmol_1} and Fig. \ref{fig:nmol_2}, and the ortho resonance of Fig. \ref{fig:nmol_3}, respectively.
For large detuning, the corresponding cavity transmission spectra show a single peak that coincides with $\omega_\textrm{c}$.
In case of (near) resonance, multiple peaks appear in the spectrum and spectral patterns can be readily interpreted along the lines described in this section.
It is obvious that panels (a) and (b) of Fig. \ref{fig:dispersion} exhibit qualitatively different behavior.
While panel (a) clearly features a characteristic three-peak structure indicating the formation of collective ortho-para polaritons, panel (b) shows two peaks implying that only the ortho molecules are involved in polariton formation.

Finally, we summarize the experimental signatures of nuclear spin isomerism in cavity transmission spectra.
For close-lying pairs of an ortho and a para rovibrational transition, with increasing density, we predict evolution from a single cavity peak to two polariton peaks, then an intermediate three-peak structure, and finally two outer peaks as the middle peak disappears.
Sweeping the cavity detuning reveals avoided crossings associated with both transitions.
On the other hand, a well-isolated ortho or para transition produces predominantly ortho or para rovibrational polaritons.
A further test proving that the three-peak structure is caused by nuclear-spin effects would be to vary the ortho/para ratio at fixed densities.
We note that the evolution from one to two, three, and finally two peaks can occur generally when the cavity couples to two nearby molecular transitions.
In our case, the two close-lying transitions belong to different spin isomers.

\section{Summary and conclusions}

In summary, we have investigated the quantum dynamics and spectroscopy of molecules coupled to an IR cavity and identified striking effects caused by the Pauli principle and nuclear spin isomerism.
First, accurate time-dependent quantum-dynamical results have been presented for two $^{14}$NH$_3$ molecules in an IR plasmonic cavity.
As a next step, the description has been generalized to molecular ensembles consisting of two distinct nuclear spin isomers in a Fabry-P\'erot cavity with experimentally feasible cavity parameters.
Using a quantum-mechanical approach, we have simulated the cavity transmission spectrum of a Fabry-P\'erot cavity coupled to an ortho-para ensemble of $^{14}$NH$_3$ molecules.
In both cases, we have found that the ortho and para spin isomers of $^{14}$NH$_3$ can form collective ortho/para molecular polaritons when both spin isomers have dipole-allowed and Pauli-allowed transitions (near) resonant with the cavity frequency.
In contrast, if only one spin isomer can couple to the cavity, the collective coupling strength is reduced accordingly and the other spin isomer is not involved in the formation of polaritons.
These findings, instead of being limited to $^{14}$NH$_3$, are generally relevant for molecules containing identical nuclei.

Our results clearly demonstrate that collective coupling to the cavity is heavily influenced by the Pauli principle and the existence of nuclear spin isomers.
The major physical insight provided by our work is that spin isomers behave as separate molecules from a polaritonic perspective and the number of molecules involved in collective coupling can differ substantially from the total number of molecules.
Since samples of molecules with identical nuclei typically contain distinct spin isomers, our results are also relevant from an experimental point of view.
In this regard, it is worth noting the light-induced drift technique \cite{99ChHe} applicable for spin isomer enrichment in experiments.
It would be of interest to test the ideas presented in this study for other molecules and experimentally verify the formation of collective ortho-para polaritons (or more generally, polaritons composed of multiple nuclear spin isomers) demonstrated theoretically in this work.
We believe that $^{14}$NH$_3$ is an ideal candidate for such experimental effort due to its small size and well-characterized high-resolution rotational-vibrational spectrum.

Beyond their fundamental significance, effects highlighted in this paper can open new perspectives for applications. 
In particular, the interplay between cavity confinement and exchange symmetry suggests the possibility of nuclear spin isomer separation and controlled nuclear spin conversion in IR cavities.
We stress that nuclear spin is conserved by the Hamiltonian used in this study and additional interactions (such as the hyperfine interaction, see for example Ref. \citenum{99ChHe}) need to be included in the Hamiltonian to enable nuclear spin conversion.
These prospects indicate that nuclear spin degrees of freedom, long recognized as essential in molecular spectroscopy, may become equally important in polaritonic chemistry.

\begin{acknowledgments}
The authors are indebted to NKFIH for funding (Grant No. K146096).
This paper was supported by the J\'anos Bolyai Research Scholarship of the Hungarian Academy of Sciences and the University of Debrecen Program for Scientific Publication.
\end{acknowledgments}

\bibliography{pauli_jctc}

\clearpage

\begin{center}
\textbf{\Large{}Supporting Information}{\Large\par}
\end{center}

\setcounter{section}{0}
\renewcommand{\thesection}{S\Roman{section}}
\setcounter{equation}{0}
\renewcommand{\theequation}{S\arabic{equation}}
\setcounter{table}{0}
\renewcommand{\thetable}{S\Roman{table}}

\section{Cavity-molecule Hamiltonian}
\label{si:a}

Let us start with the Hamiltonian of 
Eq. (1) of the manuscript.
We assume that $n$ molecules are coupled to a cavity mode, intermolecular interactions are negligible (the density is sufficiently low), the cavity couples to rovibrational transitions in the electronic ground state and the response of electrons to the external field is treated perturbatively.
The perturbation operator corresponds to
\begin{equation}
    \hat{W} = - g \sum_{i=1}^n \hat{\vec{\mu}}^{(i)} \vec{e} (\hat{a}^\dagger + \hat{a}) +
            \frac{g^2}{\hbar \omega_\textrm{c}} \left( \sum_{i=1}^n \hat{\vec{\mu}}^{(i)} \vec{e} \right)^2
\end{equation}
while the zeroth-order state is chosen as the product
\begin{equation}
    | \Phi_0 \rangle = \prod_{i=1}^{n} |\varphi_0^{(i)} \rangle
\end{equation}
of molecular electronic ground states $|\varphi_0^{(i)} \rangle$. 
Since electronic clouds of different molecules do not overlap, no antisymmetrization of $| \Phi_0 \rangle$ is required for the exchange of any two electrons between different molecules when deriving matrix elements of $\hat{W}$.

The first-order correction term due to light-matter interaction reads
\begin{equation}
    \langle \Phi_0 | \hat{W} | \Phi_0 \rangle = - g \sum_{i=1}^n \vec{\mu}_0^{(i)} \vec{e} (\hat{a}^\dagger + \hat{a}) + 
    \frac{g^2}{\hbar \omega_\textrm{c}} \Biggl[ \sum_{i=1}^n \vec{e} d_0^{(i)} \vec{e} + \sum_{i=1}^n \sum_{j \ne i}^n \left( \vec{\mu}_0^{(i)} \vec{e} \right) \left( \vec{\mu}_0^{(j)} \vec{e} \right) \Biggr]
  \label{eq:pt1}
\end{equation}
where the ground-state dipole moment equals
$\vec{\mu}_0^{(i)} = \langle \varphi_0^{(i)} | \hat{\vec{\mu}}^{(i)} | \varphi_0^{(i)} \rangle$
and components of the ground-state dipole self-energy (DSE) tensor are defined as
$(d_0)^{(i)}_{jk} = \langle \varphi_0^{(i)} | \hat{\mu}_j^{(i)} \hat{\mu}_k^{(i)} | \varphi_0^{(i)} \rangle$
with $j,k=1,2,3$.
For the sake of consistency, we also take into account the second-order correction term 
\begin{equation}
    \sum_{k>0} \frac{|\langle \Phi_0 | \hat{W} | \Phi_k \rangle |^2}{E_0-E_k} \approx 
    g^2 \sum_{k>0} \frac{|\langle \Phi_0 | \sum_{i=1}^n \hat{\vec{\mu}}^{(i)} \vec{e} | \Phi_k \rangle |^2}{E_0-E_k} (\hat{a}^\dagger + \hat{a})^2
  \label{eq:pt2}
\end{equation}
where only the dipole interaction is retained in $\hat{W}$ (terms higher in order than $g^2$ are neglected) and zeroth-order electronic energy levels and states are denoted by $E_k$ and $| \Phi_k \rangle$, respectively.
As the operators $\hat{\vec{\mu}}^{(i)}$ act on a single molecule, the summation in Eq. \eqref{eq:pt2} can be restricted to zeroth-order electronic states
\begin{equation}
    | \Phi_k^{(j)} \rangle = |\varphi_k^{(j)} \rangle \prod_{i=1,i \ne j}^{n} |\varphi_0^{(i)} \rangle
\end{equation}
and zeroth-order energy levels
$E_k^{(j)} = \epsilon_k^{(j)} + \sum_{i=1,i\ne j}^{n} \epsilon_0^{(i)}$.
Here, molecule $j$ is excited to electronic state $|\varphi_k^{(j)} \rangle$ (with electronic energy $\epsilon_k^{(j)}$) and the remaining molecules occupy their respective electronic ground states.
With this, the second-order correction can be expressed as
\begin{gather}
    E_0^{(2)} =  
    g^2 \sum_{k>0} \frac{|\langle \Phi_0 | \sum_{i=1}^n \hat{\vec{\mu}}^{(i)} \vec{e} | \Phi_k \rangle |^2}{E_0-E_k} (\hat{a}^\dagger + \hat{a})^2 \nonumber \\
    = g^2 \sum_{i=1}^n \sum_{k>0} \frac{|\langle \varphi_0^{(i)} | \hat{\vec{\mu}}^{(i)} \vec{e} | \varphi_k^{(i)} \rangle |^2}{\epsilon_0^{(i)}-\epsilon_k^{(i)}} (\hat{a}^\dagger + \hat{a})^2 =
    - \frac{g^2}{2} \sum_{i=1}^n \vec{e} \alpha_0^{(i)} \vec{e} (\hat{a}^\dagger + \hat{a})^2
\end{gather}
where $\alpha_0^{(i)}$ is the ground-state polarizability of molecule $i$.
This way, all light-matter interaction terms up to $g^2$ are included in the working Hamiltonian of 
Eq. (2) of the manuscript.

\section{Tavis--Cummings model}
\label{si:b}

The secular equation of $\mathbf{H}$ defined in 
Eq. (4) of the manuscript takes the form
\begin{gather}
    \sum_{i=1}^{n_\textrm{o}} V_\textrm{o}^{(i)} x_i + \sum_{j=1}^{n_\textrm{p}} V_\textrm{p}^{(j)} y_j = \lambda b \nonumber \\
    V_\textrm{o}^{(i)} b + \Delta_\textrm{o} x_i = \lambda x_i  \textrm{,} ~~~ i = 1,\dots,n_\textrm{o} \label{eq:secular} \\
    V_\textrm{p}^{(j)} b + \Delta_\textrm{p} y_j = \lambda y_j \textrm{,} ~~~ j = 1,\dots,n_\textrm{p} \nonumber
\end{gather}
for $n = n_\textrm{o}+n_\textrm{p}$ molecules, where $\lambda$ and $\mathbf{v} = (b,x_1,\dots,x_{n_\textrm{o}},y_1,\dots,y_{n_\textrm{p}})^\textrm{T}$ refer to the eigenvalues and eigenvectors of $\mathbf{H}$.
The secular equation of Eq. \eqref{eq:secular} is amenable to an analytical solution.
It is straightforward to see that there is an $(n_\textrm{o}-1)$-fold degenerate eigenvalue $\lambda = \Delta_\textrm{o}$ with $n_\textrm{o}-1$ eigenvectors satisfying 
$b = 0$, $y_j=0$ for $j=1,\dots,n_\textrm{p}$, and
$\sum_{i=1}^{n_\textrm{o}} V_\textrm{o}^{(i)} x_i = 0$.
Similarly, for $\lambda = \Delta_\textrm{p}$, there are $n_\textrm{p}-1$ eigenvectors defined by
$b = 0$, $x_i=0$ for $i=1,\dots,n_\textrm{o}$, and
$\sum_{j=1}^{n_\textrm{p}} V_\textrm{p}^{(j)} y_j = 0$.
In total, these solutions correspond to $n_\textrm{o}+n_\textrm{p}-2$ dark states that have zero overlap with Fock state $|1\rangle$.
Orthonormal sets of ortho and para dark states can be constructed by the Gram--Schmidt process which yields
\begin{equation}
    \mathbf{x}^{(i)} = 
    \frac{|V_\textrm{o}^{(i+1)}|}{\sqrt{\sum_{j=1}^i |V_\textrm{o}^{(j)}|^2 \sum_{j=1}^{i+1} |V_\textrm{o}^{(j)}|^2}}
    (V_\textrm{o}^{(1)},\dots,V_\textrm{o}^{(i)},-\frac{\sum_{j=1}^i |V_\textrm{o}^{(j)}|^2}{V_\textrm{o}^{(i+1)}},0,\dots)^\textrm{T}
    \label{eq:ortho_dark}
\end{equation}
for $i=1,\dots,n_\textrm{o}-1$ and
\begin{equation}
    \mathbf{y}^{(j)} = 
    \frac{|V_\textrm{p}^{(j+1)}|}{\sqrt{\sum_{k=1}^j |V_\textrm{p}^{(k)}|^2 \sum_{k=1}^{j+1} |V_\textrm{p}^{(k)}|^2}}
    (V_\textrm{p}^{(1)},\dots,V_\textrm{p}^{(j)},-\frac{\sum_{k=1}^j |V_\textrm{p}^{(k)}|^2}{V_\textrm{p}^{(j+1)}},0,\dots)^\textrm{T}
    \label{eq:para_dark}
\end{equation}
for $j=1,\dots,n_\textrm{p}-1$.
The remaining three eigenstates are bright polaritonic states for which $\lambda \ne \Delta_\textrm{o}$ and $\lambda \ne \Delta_\textrm{p}$.
From Eq. \eqref{eq:secular}, eigenvector coefficients other than $b$ can be expressed in terms of $b$ as
\begin{equation}
    x_i = \frac{V_\textrm{o}^{(i)}}{\lambda-\Delta_\textrm{o}} b 
    \label{eq:bright1}
\end{equation}
for $i = 1,\dots,n_\textrm{o}$ and
\begin{equation}
    y_j = \frac{V_\textrm{p}^{(j)}}{\lambda-\Delta_\textrm{p}} b
    \label{eq:bright2}
\end{equation}
for $j = 1,\dots,n_\textrm{p}$.
It is clear from Eq. \eqref{eq:bright1} and Eq. \eqref{eq:bright2} that $x_i \sim V_\textrm{o}^{(i)}$ and $y_j \sim V_\textrm{p}^{(j)}$ for the three bright states.
This motivates the definition of the three bright basis vectors
\begin{gather}
    \mathbf{w}_0 = (1,0,\dots,0,0,\dots,0)^\textrm{T} \nonumber \\
    \mathbf{w}_1 = \frac{1}{v_\textrm{o}}
        (0,V_\textrm{o}^{(1)},\dots,V_\textrm{o}^{(n_\textrm{o})},0,\dots,0)^\textrm{T} \\
    \mathbf{w}_2 = \frac{1}{v_\textrm{p}}
         (0,0,\dots,0,V_\textrm{p}^{(1)},\dots,V_\textrm{p}^{(n_\textrm{p})})^\textrm{T} \nonumber
\end{gather}
where 
$v_\textrm{o} = \sqrt{\sum_{i=1}^{n_\textrm{o}}|V_\textrm{o}^{(i)}|^2}$
and
$v_\textrm{p} = \sqrt{\sum_{j=1}^{n_\textrm{p}}|V_\textrm{p}^{(j)}|^2}$.
The vectors $\mathbf{w}_0$, $\mathbf{w}_1$ and $\mathbf{w}_2$ are orthogonal to the dark eigenvectors of $\mathbf{H}$.
The Hamiltonian $\mathbf{H}$ can be transformed to the basis spanned by $\mathbf{w}_i$, which gives the three-dimensional bright Hamiltonian
\begin{equation}
    \mathbf{H}_\textrm{bright} = 
    \left[
    \begin{array}{ccc}
            0 & v_\textrm{o} & v_\textrm{p} \\
            v_\textrm{o} & \Delta_\textrm{o} & 0 \\
            v_\textrm{p} & 0 & \Delta_\textrm{p} 
        \end{array}
    \right].
\end{equation}
The characteristic equation of $\mathbf{H}_\textrm{bright}$ reads
\begin{equation}
    \lambda^3 - (\Delta_\textrm{o}+\Delta_\textrm{p}) \lambda^2 +
    (\Delta_\textrm{o}\Delta_\textrm{p} - v_\textrm{o}^2 - v_\textrm{p}^2) \lambda +
    (v_\textrm{o}^2 \Delta_\textrm{p} + v_\textrm{p}^2 \Delta_\textrm{o}) = 0
\end{equation}
which can be solved for the energies of the bright states.
For $\Delta_\textrm{o}=\Delta_\textrm{p}=\Delta$, it is easy to see that $\lambda_0 = \Delta$ is an eigenvalue of $\mathbf{H}_\textrm{bright}$ and the remaining two eigenvalues satisfy
\begin{equation}
    \lambda^2 - \Delta \lambda - (v_\textrm{o}^2 + v_\textrm{p}^2) = 0
\end{equation}
which yields 
\begin{equation}
    \lambda_\pm = \frac{\Delta \pm \sqrt{\Delta^2+4(v_\textrm{o}^2 + v_\textrm{p}^2)}}{2}.
\end{equation}
In this special case, the eigenvectors of $\mathbf{H}_\textrm{bright}$ can be expressed as
\begin{gather}
    \mathbf{v}_0 = \frac{1}{\sqrt{v_\textrm{o}^2 + v_\textrm{p}^2}} (0,v_\textrm{p},-v_\textrm{o})^\textrm{T} \nonumber \\
    \mathbf{v}_\pm = \frac{1}{\sqrt{(\lambda_\pm - \Delta)^2 + v_\textrm{o}^2 + v_\textrm{p}^2}} 
    (\lambda_\pm -\Delta,v_\textrm{o},v_\textrm{p})^\textrm{T}.
    \label{eq:eigvec_special}
\end{gather}
Eq. \eqref{eq:eigvec_special} implies that $\mathbf{v}_0$ becomes a dark state ($\mathbf{w}_0^\textrm{T} \mathbf{v}_0 = 0$) for $\Delta_\textrm{o}=\Delta_\textrm{p}$.

\section{Light-matter interaction matrix elements}
\label{si:c}

The starting point is the definition of symmetric-top eigenfunctions
\begin{equation}
    |JKM\rangle = \sqrt{\frac{2J+1}{8\pi^2}} D^{(J)*}_{MK} (\phi,\theta,\chi)
\label{eq:jkmd}
\end{equation}
where $D^{(J)}_{MK} (\phi,\theta,\chi)$ denotes elements of the Wigner D matrix, and $\phi$, $\theta$ and $\chi$ are
the Euler angles.\cite{06BuJe}
In the following derivations, we will rely on the concept of tensor operators.\cite{17SaNa}
Matrix elements of the electric dipole moment can be derived in the $|v^pJKM\rangle$ eigenbasis as follows.
First, Cartesian space-fixed (SF) components of the dipole moment are transformed to the so-called spherical basis, that is,
\begin{gather}
    \mu^\textrm{SF}_x = \frac{1}{\sqrt{2}} (\mu^\textrm{SF}_{-1} - \mu^\textrm{SF}_{+1}) \nonumber \\
    \mu^\textrm{SF}_y = \frac{\textrm{i}}{\sqrt{2}} (\mu^\textrm{SF}_{-1} + \mu^\textrm{SF}_{+1}) \label{eq:spherical} \\
    \mu^\textrm{SF}_z = \mu^\textrm{SF}_{0}. \nonumber
\end{gather}
Then, the SF components are expressed in terms of the body-fixed (BF) components as
\begin{equation}
    \mu^\textrm{SF}_\alpha = \sum_{\beta=-1}^{+1} D^{(1)*}_{\alpha\beta} (\phi,\theta,\chi) \mu^\textrm{BF}_\beta
\label{eq:rotation}
\end{equation}
with $\alpha = 0,\pm1$.
The reason for working with the spherical basis is that the BF components of the dipole can be rotated to the SF frame by the $J=1$ Wigner D matrix (in other words, the dipole moment expressed in the spherical basis is a spherical tensor operator of rank one).
Eq. \eqref{eq:rotation} is useful when matrix elements of the SF dipole moment are evaluated in the $|v^pJKM\rangle$ eigenbasis, which reads
\begin{gather}
    \langle v^pJKM| \mu^\textrm{SF}_\alpha |v'^{p'}J'K'M'\rangle = \nonumber
    \sum_{\beta=-1}^{+1} \langle v^pJKM| D^{(1)*}_{\alpha\beta} \mu^\textrm{BF}_\beta |v'^{p'}J'K'M'\rangle \nonumber \\
    = (-1)^{M-K} \sqrt{(2J+1)(2J'+1)} 
    \begin{pmatrix}
        J & 1 & J'\\
       -M & \alpha & M'
    \end{pmatrix}
    \sum_{\beta=-1}^{+1} \langle v^{p} | \mu^\textrm{BF}_\beta | v'^{p'}\rangle
    \begin{pmatrix}
        J & 1 & J'\\
       -K & \beta & K'
    \end{pmatrix} \label{eq:dipmx_app} \\
    = (-1)^{M-K} \sqrt{(2J+1)(2J'+1)} \langle v^{p} | \mu^\textrm{BF}_0 | v'^{p'} \rangle
    \begin{pmatrix}
        J & 1 & J'\\
       -M & \alpha & M'
    \end{pmatrix} 
    \begin{pmatrix}
        J & 1 & J'\\
       -K & 0 & K'
    \end{pmatrix} \nonumber
\end{gather}
where Wigner's 3-j symbol is denoted by
\[
\begin{pmatrix}
    j_1 & j_2 & j_3\\
    m_1 & m_2 & m_3
\end{pmatrix}.
\]
In the last line of Eq. \eqref{eq:dipmx_app}, we have exploited that only the $\langle v^{p} | \mu^\textrm{BF}_0 | v'^{p'}\rangle = \langle v^{p} | \mu^\textrm{BF}_z | v'^{p'}\rangle$ component of the BF dipole matrix is non-zero.
The 3-j symbols vanish unless conditions $|j_1-j_2| \le j_3 \le j_1+j_2$ and $m_1 + m_2 + m_3 = 0$ are satisfied. 
These relationships can be used to determine the selection rules for dipole transitions in this particular case, that is, $|J-1| \le J' \le J+1$, $M = M'+\alpha$ and $K=K'$.

Since the DSE and polarizability tensors are symmetric tensors of rank two, we introduce a common notation $T$ for them.
Spherical tensor components of a rank-two symmetric tensor $T$ are defined as
\begin{gather}
    T^{(2)}_{\pm2} = \frac{1}{2}(T_{xx}-T_{yy}) \pm \textrm{i} T_{xy} \nonumber \\
    T^{(2)}_{\pm1} = \mp (T_{xz} \pm \textrm{i} T_{yz}) \\
    T^{(2)}_{0} = \frac{1}{\sqrt{6}} (2T_{zz}-T_{xx}-T_{yy}) \nonumber
\end{gather}
for $J=2$
and
\begin{equation}
    T^{(0)}_{0} = -\frac{1}{\sqrt{3}} (T_{xx}+T_{yy}+T_{zz})
\end{equation}
for $J=0$, while the quantities $T^{(1)}_\alpha$ ($\alpha=0, \pm 1$, for $J=1$) are identically zero.
Next, we move to the BF frame and exploit that
$T^\textrm{BF}_{xx} = T^\textrm{BF}_{yy} \ne 0$, 
$T^\textrm{BF}_{zz} \ne 0$ and 
$T^\textrm{BF}_{xy} = T^\textrm{BF}_{xz} = T^\textrm{BF}_{yz} = 0$
in our computational model. These relations yield
\begin{gather}
    T^{(2),\textrm{BF}}_{\pm2} = 0 \nonumber \\
    T^{(2),\textrm{BF}}_{\pm1} = 0 \\
    T^{(2),\textrm{BF}}_{0} = \sqrt{\frac{2}{3}} (T^\textrm{BF}_{zz}-T^\textrm{BF}_{xx}) \nonumber
\end{gather}
and
\begin{equation}
    T^{(0),\textrm{BF}}_{0} = -\frac{1}{\sqrt{3}} (2T^\textrm{BF}_{xx}+T^\textrm{BF}_{zz}).
\end{equation}
If the cavity polarization vector is chosen as $\mathbf{e} = (0,0,1)$, only the $T^\textrm{SF}_{zz}$ tensor component appears in the Hamiltonian.
Using the transformation property
\begin{equation}
    T^{(k),\textrm{SF}}_{\alpha} = \sum_{\beta=-k}^k D^{(k)*}_{\alpha\beta} T^{(k),\textrm{BF}}_{\beta}
\end{equation}
of spherical tensors under rotation, $T^\textrm{SF}_{zz}$ can be expressed in terms of BF spherical tensors, that is,
\begin{gather}
    T^\textrm{SF}_{zz} = -\frac{1}{\sqrt{3}} T^{(0),\textrm{SF}}_{0} +\sqrt{\frac{2}{3}} T^{(2),\textrm{SF}}_{0} 
    = 
    -\frac{1}{\sqrt{3}} D^{(0)*}_{00} T^{(0),\textrm{BF}}_{0} + \sqrt{\frac{2}{3}} \sum_{\beta=-2}^2 D^{(2)*}_{0\beta} T^{(2),\textrm{BF}}_{\beta} \nonumber \\
    = \frac{2T^\textrm{BF}_{xx}+T^\textrm{BF}_{zz}}{3} D^{(0)*}_{00} + \frac{2(T^\textrm{BF}_{zz}-T^\textrm{BF}_{xx})}{3} D^{(2)*}_{00}.
\end{gather}
With this, matrix elements of $T^\textrm{SF}_{zz}$ in the $|v^{p}JKM\rangle$ eigenbasis read
\begin{gather}
    \langle v^{p}JKM| T^\textrm{SF}_{zz} |v'^{p'}J'K'M'\rangle = 
    \langle v^{p}JKM| \frac{2T^\textrm{BF}_{xx}+T^\textrm{BF}_{zz}}{3} D^{(0)*}_{00} |v'^{p'}J'K'M'\rangle \nonumber \\ 
    + \langle v^{p}JKM| \frac{2(T^\textrm{BF}_{zz}-T^\textrm{BF}_{xx})}{3} D^{(2)*}_{00} |v'^{p'}J'K'M'\rangle = (-1)^{M-K} \sqrt{(2J+1)(2J'+1)} \label{eq:dsemx_app} \\
     \left[ a_{v^{p}v'^{p'}}
    \begin{pmatrix}
        J & 0 & J'\\
       -M & 0 & M'
    \end{pmatrix}
    \begin{pmatrix}
        J & 0 & J'\\
       -K & 0 & K'
    \end{pmatrix}+
    b_{v^{p}v'^{p'}}
    \begin{pmatrix}
        J & 2 & J'\\
       -M & 0 & M'
    \end{pmatrix}
    \begin{pmatrix}
        J & 2 & J'\\
       -K & 0 & K'
    \end{pmatrix} \right] \nonumber
\end{gather}
which simplifies to
\begin{gather}
    \langle v^{p}JKM| T^\textrm{SF}_{zz} |v'^{p'}J'K'M'\rangle 
        = a_{v^{p}v'^{p'}} \delta_{JJ'} \delta_{KK'} \delta_{MM'} +
    (-1)^{M-K} \sqrt{(2J+1)(2J'+1)} \nonumber \\
    b_{v^{p}v'^{p'}}
    \begin{pmatrix}
        J & 2 & J'\\
       -M & 0 & M'
    \end{pmatrix}
    \begin{pmatrix}
        J & 2 & J'\\
       -K & 0 & K'
    \end{pmatrix}
    \nonumber
\end{gather}
where 
\begin{gather}
a_{v^{p}v'^{p'}}=\langle v^{p} | \frac{2T^\textrm{BF}_{xx}+T^\textrm{BF}_{zz}}{3} | v'^{p'} \rangle
\nonumber \\
b_{v^{p}v'^{p'}}=\langle v^{p} | \frac{2(T^\textrm{BF}_{zz}-T^\textrm{BF}_{xx})}{3} | v'^{p'} \rangle
\nonumber
\end{gather}
and Wigner D matrix elements
$\langle JKM| D^{(k)*}_{\alpha\beta} |J'K'M'\rangle$
are evaluated analytically (see also Eq. \eqref{eq:dipmx_app}). The 3-j symbols in Eq. \eqref{eq:dsemx_app} 
imply that selection rules $M=M'$ and $K=K'$ hold for $\mathbf{e} = (0,0,1)$.

\section{Symmetry data}
\label{si:d}

The character table of $D_{3\textrm{h}}(\textrm{M})$ is given in Table \ref{tbl:d3h} (see also Table A-10 of Ref. \onlinecite{06BuJe}).
Spin eigenfunctions of three identical spin-1/2 particles are listed in Table \ref{tbl:spin} together with nuclear spin quantum numbers ($I$ and $M_I=-I,\dots,I$) and irreducible representation labels ($\Gamma_\textrm{spin}$) of the $D_{3\textrm{h}}(\textrm{M})$ molecular symmetry group.

\begin{table}[h]
  \caption{\label{tbl:d3h}
  Character table of the $D_{3\textrm{h}}(\textrm{M})$ molecular symmetry group.}
  \begin{tabular}{c|c|c|c|c|c|c}
          & ~~ $E$ ~~ & ~ $(123)/(132)$ ~ & ~ $(12)/(13)/(23)$ & ~~ $E^*$ ~~ & ~ $(123)^*/(132)^*$ ~ & ~ $(12)^*/(13)^*/(23)^*$ ~ \\
        \hline
    $\textrm{A}_1'$ ~ & $1$ & $1$ & $1$ & $1$ & $1$ & $1$ \\
    $\textrm{A}_1''$ ~ & $1$ & $1$ & $1$ & $-1$ & $-1$ & $-1$ \\
    $\textrm{A}_2'$ ~ & $1$ & $1$ & $-1$ & $1$ & $1$ & $-1$ \\
    $\textrm{A}_2''$ ~ & $1$ & $1$ & $-1$ & $-1$ & $-1$ & $1$ \\
    $\textrm{E}'$ ~ & $2$ & $-1$ & $0$ & $2$ & $-1$ & $0$ \\
    $\textrm{E}''$ ~ & $2$ & $-1$ & $0$ & $-2$ & $1$ & $0$ \\
  \end{tabular}
\end{table}

\begin{table}[h]
  \caption{\label{tbl:spin}
  Spin eigenfunctions of three spin-1/2 particles (protons in this case), nuclear spin quantum numbers ($I$ and $M_I$) and symmetry properties ($\Gamma_\textrm{spin}$, irreducible representations of the $D_{3\textrm{h}}(\textrm{M})$ molecular symmetry group).}
  \begin{tabular}{c|c|c|c}
        spin function  & ~~ $I$ ~~ & ~ $M_I$ ~ & ~ $\Gamma_\textrm{spin}$ ~ \\
        \hline
    $|\alpha\alpha\alpha\rangle$ & $3/2$ & $3/2$ & $\textrm{A}_1'$ \\
    $\frac{1}{\sqrt{3}}(|\alpha\alpha\beta\rangle+|\alpha\beta\alpha\rangle+|\beta\alpha\alpha\rangle)$ & $3/2$ & $1/2$ & $\textrm{A}_1'$ \\
    $\frac{1}{\sqrt{3}}(|\beta\beta\alpha\rangle+|\beta\alpha\beta\rangle+|\alpha\beta\beta\rangle)$ & $3/2$ & $-1/2$ & $\textrm{A}_1'$ \\
    $|\beta\beta\beta\rangle$ & $3/2$ & $-3/2$ & $\textrm{A}_1'$ \\
    \hline
    $\frac{1}{\sqrt{6}}(2|\alpha\alpha\beta\rangle-|\alpha\beta\alpha\rangle-|\beta\alpha\alpha\rangle)$,
    $\frac{1}{\sqrt{2}} (|\alpha\beta\alpha\rangle-|\beta\alpha\alpha\rangle)$
    & $1/2$ & $1/2$ & $\textrm{E}'$ \\
    $\frac{1}{\sqrt{6}}(2|\beta\beta\alpha\rangle-|\beta\alpha\beta\rangle-|\alpha\beta\beta\rangle)$,
    $\frac{1}{\sqrt{2}} (|\beta\alpha\beta\rangle-|\alpha\beta\beta\rangle)$
    & $1/2$ & $-1/2$ & $\textrm{E}'$
  \end{tabular}
\end{table}

\section{Electronic structure data}
\label{si:e}

\begin{table}[h]
  \caption{\label{tbl:elstruc}
    Relevant electronic structure data (CCSD/aug-cc-pVTZ level of theory) for the effective one-dimensional inversion model of $^{14}$NH$_3$. Each quantity (see the manuscript for notations) is given in atomic units.}
  \scriptsize
  \begin{tabular}{c|c|c|c|c|c}
        ~~ $x ~/~ \textrm{au}$ ~~ & ~~ $\mu^\textrm{BF}_z ~/~ \textrm{au}$ ~~ & ~~ $\alpha^\textrm{BF}_{xx} ~/~ \textrm{au}$ ~~ & ~~ $\alpha^\textrm{BF}_{zz} ~/~ \textrm{au}$ ~~ & ~~ $d^\textrm{BF}_{xx} ~/~ \textrm{au}$ ~~ & ~~ $d^\textrm{BF}_{zz} ~/~ \textrm{au}$ ~~ \\
        \hline
        $-2.003410$ & $0.936633$ & $15.324192$ & $17.823569$ & $4.897186$ & $6.631130$ \\
        $-1.836460$ & $0.940011$ & $14.493735$ & $16.608771$ & $4.804047$ & $6.346852$ \\
        $-1.669509$ & $0.926242$ & $13.996242$ & $15.755016$ & $4.752142$ & $6.090999$ \\
        $-1.502558$ & $0.897383$ & $13.696469$ & $15.147945$ & $4.725807$ & $5.852953$ \\
        $-1.335607$ & $0.855513$ & $13.523391$ & $14.740748$ & $4.716123$ & $5.630835$ \\
        $-1.168656$ & $0.801357$ & $13.430496$ & $14.514361$ & $4.716692$ & $5.425640$ \\
        $-1.001705$ & $0.734311$ & $13.384913$ & $14.459044$ & $4.722841$ & $5.239388$ \\
        $-0.834754$ & $0.652733$ & $13.359492$ & $14.561628$ & $4.730822$ & $5.074029$ \\
        $-0.667803$ & $0.554611$ & $13.332448$ & $14.794823$ & $4.737831$ & $4.931610$ \\
        $-0.500853$ & $0.438499$ & $13.291411$ & $15.109220$ & $4.742306$ & $4.814692$ \\
        $-0.333902$ & $0.304730$ & $13.237812$ & $15.429222$ & $4.744003$ & $4.726602$ \\
        $-0.166951$ & $0.156490$ & $13.189130$ & $15.669426$ & $4.743955$ & $4.671295$ \\
        $0.000000$ & $0.000000$ & $13.169172$ & $15.757708$ & $4.743688$ & $4.652358$ \\
        $0.166951$ & $-0.156490$ & $13.189130$ & $15.669426$ & $4.743955$ & $4.671295$ \\
        $0.333902$ & $-0.304730$ & $13.237812$ & $15.429222$ & $4.744003$ & $4.726602$ \\
        $0.500853$ & $-0.438499$ & $13.291411$ & $15.109220$ & $4.742306$ & $4.814692$ \\
        $0.667803$ & $-0.554611$ & $13.332448$ & $14.794823$ & $4.737831$ & $4.931610$ \\
        $0.834754$ & $-0.652733$ & $13.359492$ & $14.561628$ & $4.730822$ & $5.074029$ \\
        $1.001705$ & $-0.734311$ & $13.384913$ & $14.459044$ & $4.722841$ & $5.239388$ \\
        $1.168656$ & $-0.801357$ & $13.430496$ & $14.514361$ & $4.716692$ & $5.425640$ \\
        $1.335607$ & $-0.855513$ & $13.523391$ & $14.740748$ & $4.716123$ & $5.630835$ \\
        $1.502558$ & $-0.897383$ & $13.696469$ & $15.147945$ & $4.725807$ & $5.852953$ \\
        $1.669509$ & $-0.926242$ & $13.996242$ & $15.755016$ & $4.752142$ & $6.090999$ \\
        $1.836460$ & $-0.940011$ & $14.493735$ & $16.608771$ & $4.804047$ & $6.346852$ \\
        $2.003410$ & $-0.936633$ & $15.324192$ & $17.823569$ & $4.897186$ & $6.631130$ \\
  \end{tabular}
\end{table}

\end{document}